\PassOptionsToPackage{dvipsnames}{xcolor}
\documentclass[a4paper,twocolumn,superscriptaddress,preprintnumbers,nobalancelastpage,longbibliography,unpublished]{quantumarticle}
\pdfoutput=1
\usepackage[utf8]{inputenc}
\usepackage[english]{babel}
\usepackage[T1]{fontenc}
\usepackage{amsmath}
\usepackage{amsfonts}
\usepackage{amsthm}
\usepackage{booktabs}
\usepackage{soul}
\usepackage[hyperindex = {true},
colorlinks = {true},
linkcolor = {blue},
citecolor = {blue}]{hyperref}
\newcommand{\MYhref}[3][blue]{\href{#2}{\color{#1}{#3}}}%
\usepackage{cleveref}
\usepackage{physics}
\usepackage[numbers,sort&compress]{natbib}
\usepackage{tikz}
\usetikzlibrary{quantikz}
\usepackage{tikzscale} 
\usepackage{subcaption}

\providecommand*{\iu}%
{\ensuremath{\mathrm{i}}}

\theoremstyle{definition}
\newtheorem{lemma}{Lemma}[section]
\newtheorem{corollary}{Corollary}[section]

\newcommand{\mcaption}[1]{\caption{{\em\small #1}}}

\begin{document}

\title{Digital quantum simulation of Schr\"{o}dinger dynamics using adaptive approximations of potential functions}

\author{Tenzan Araki}
\orcid{0000-0002-9510-4069}
\email{taraki@ethz.ch}
\affiliation{Quantum Engineering, Department of Information Technology and Electrical Engineering,  ETH Z\"{u}rich, 8092 Z\"{u}rich, Switzerland}
\author{James Stokes}
\affiliation{Department of Mathematics, University of Michigan, Ann Arbor, MI 48109 USA}
\author{Shravan Veerapaneni}
\affiliation{Department of Mathematics, University of Michigan, Ann Arbor, MI 48109 USA}
\maketitle
\newpage
\begin{abstract}
Digital quantum simulation (DQS) of continuous-variable quantum systems in the position basis requires efficient implementation of diagonal unitaries approximating the time evolution operator generated by the potential energy function. In this work, we provide efficient implementations suitable for potential functions approximable by piecewise polynomials, with either uniform or adaptively chosen subdomains. For a fixed precision of approximation, we show how adaptive grids can significantly reduce the total gate count at the cost of introducing a small number of ancillary qubits. We demonstrate the circuit construction with both physically motivated and artificially designed potential functions, and discuss their generalizations to higher dimensions.
\end{abstract}

\section{Introduction}\label{section : introduction}
A problem of fundamental importance to many areas of science and engineering is to solve the initial value problem for the $d$-dimensional many-body time-dependent Schr\"{o}dinger equation (TDSE) with potential energy function $V : \mathbb{R}^d \longrightarrow\mathbb{R}$,
\begin{align}
    & \iu \partial_t \psi(t,x)
 = \left[- \frac{1}{2} \nabla^2 + V(x)\right]\psi(t,x), & t \in (0,T) \notag, \\
 & \psi(0,\cdot) = \psi_0 \in L^2(\mathbb{R}^d). & \label{e:IVP}
\end{align}
The above formulation is known as first quantization in the real-space (position) basis. Applications include simulations of chemical dynamics \cite{kassal2008polynomial} and scattering events in bosonic quantum field theories \cite{jordan2011quantum}, among others.

\begin{figure*}[t]
  \centering
  \includegraphics[width=\textwidth]{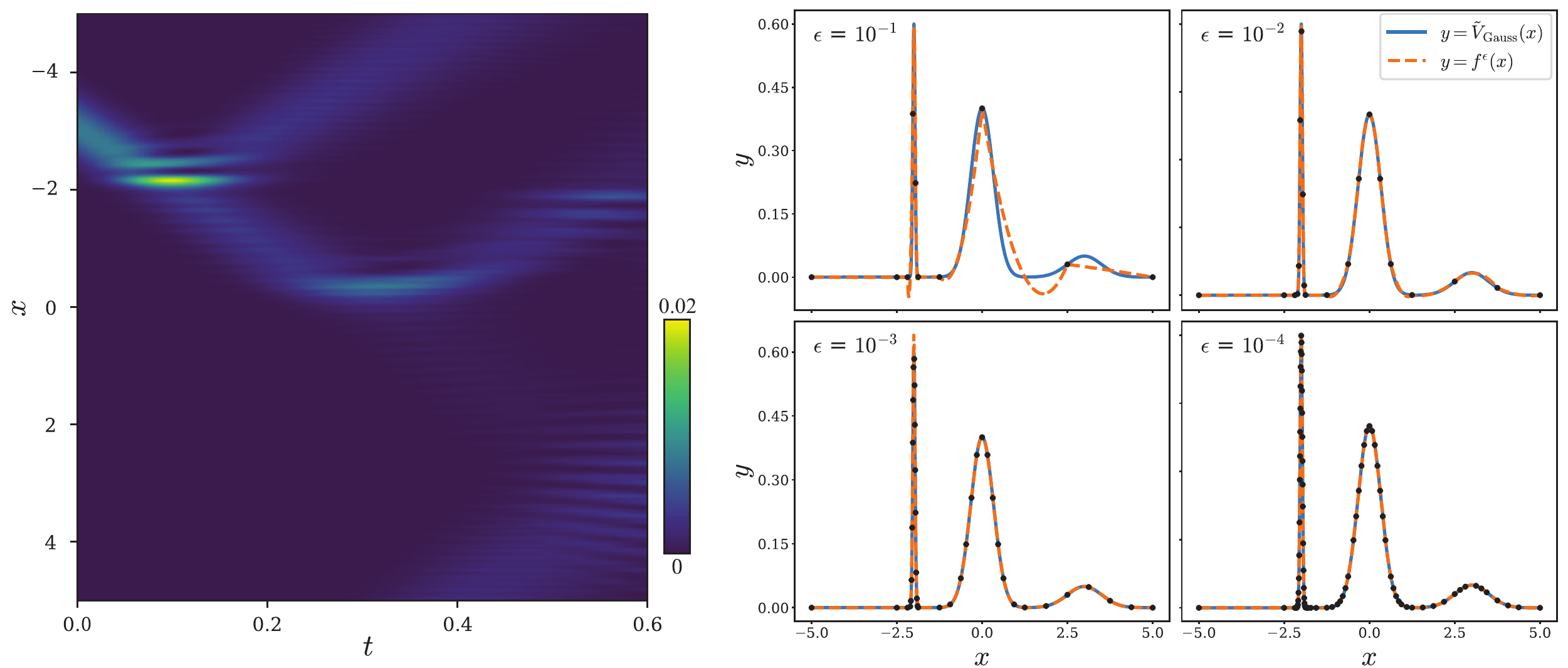}
\mcaption{An example of DQS showcasing the solution of \eqref{e:IVP} for a potential function with multi-scale features. (Left) The space-time dependence of the probability density $|\psi(x,t)|^2$ obtained from DQS of a quantum particle in a triple Gaussian potential $V_{\text{Gauss}}$. (Right) The piecewise quadratic approximation $f^{\epsilon}(x)$ of the rescaled triple Gaussian function $\tilde{V}_{\text{Gauss}}(x) = V_{\text{Gauss}}(x) \Delta t$ obtained classically for successive approximations $\epsilon \in \{10^{-1}, 10^{-2}, 10^{-3}, 10^{-4}\}$. The resulting piecewise functions have non-uniform subdomains indicated by the positions of the knots (black dots). 
  When compared to uniform grids, the number of subdomains grows more modestly as $\epsilon$ is reduced for functions with multi-scale features, such as in this potential function. In this work, we propose algorithms that encode piecewise polynomial representations on a quantum computer and demonstrate that the quantum circuits for adaptive grids have significantly lower gate counts compared to those of uniform grids for a fixed $\epsilon$.
More details on this example are provided in \Cref{subsubsection : num exp results gaussians}.}
  \label{fig: triple gaussian example}
\end{figure*}

An algorithm for solving \eqref{e:IVP} using DQS, illustrated for the one-dimensional case in \Cref{fig: qc with Z-W}, was originally put forth by Zalka and Wiesner \cite{zalka1998, wiesner1996}. Their algorithm introduces two approximations associated with time-stepping and the discretization of the spatial domain. First, the time evolution operator is replaced by alternating pulses of time evolution generated by the kinetic and potential energy operators, respectively. Second, the infinite-dimensional kinetic and potential energy generators are replaced by matrix approximations acting on finite-dimensional subspaces. Assuming for simplicity that $\psi(t,\cdot)$ is given by a periodic continuation from the fundamental domain $\Omega := [0,1)^d \subseteq \mathbb{R}^d$, the kinetic and potential operators admit simple discretizations associated with uniform tensor product grids in Fourier and position space, respectively. Zalka and Wiesner showed that the discretized kinetic time evolution operator admits an efficient implementation based on the quantum Fourier transform (QFT). In contrast, the problem of applying the diagonal unitary generated by the potential operator depends on the nature of the potential function. This work focuses on the latter problem, specifically via approximating the potential energy function by a piecewise polynomial function with either uniform or adaptively chosen subdomains. Although our motivation is to perform quantum simulation of the TDSE, the results are applicable to any situation requiring the approximation of unitary operators which are diagonal in the position basis.

Function approximation using piecewise polynomials is a common practice in classical computing (e.g., see \cite{driscoll2014chebfun}), especially for efficient computation in the case of functions that have features with widely disparate length-scales (such as in Fig. \ref{fig: triple gaussian example}). However, in the context of quantum computing, a na\"{i}ve case-distinction based on quantum measurements is not feasible due to the inevitability of collapsing the wave function. This implies that classical techniques of evaluating piecewise polynomial functions, which generally require subdomain-dependent operations, cannot be straightforwardly applied to a quantum computer.

In this work, we introduce quantum algorithms to apply diagonal unitaries generated by piecewise polynomial functions, assuming either uniform or adaptive subdomains. In the case of uniform subdomains, an ancilla-free algorithm can be implemented using only rotation-Z and CNOT gates. In problems requiring adaptive subdomains, we provide an algorithm utilizing an ancillary quantum register, which can nevertheless significantly reduce the required number of single- and two-qubit operations.

The article is organized as follows. We first review the theory behind our ancilla-free and ancilla-assisted quantum algorithms in \Cref{sec:theory}. This section introduces classical preparations in the form of \Cref{lem:general}, which provides the basis for our proposed quantum circuit constructions outlined in \Cref{section : quantum implementation}, and furthermore explains the relationship with previous work in \Cref{sec:previous}. \Cref{sec:examples} is dedicated to explaining the proposal by way of explicit quantum circuits for uniform and adaptive subdomains in one dimension, followed by their extensions to $d>1$. \Cref{section : num exp} provides numerical support for our proposal, focusing on three choices of potential function in one dimension: the cosine function (which appears in, e.g., the quantum kicked rotor model \cite{georgeot2001, levi2003}, quantum kicked Harper model \cite{levi2004}, and the Kogut-Susskind model with $U(1)$ gauge group \cite{kogut1975hamiltonian, sohaib2022}), the Eckart barrier \cite{eckart1930}, and an artificial potential with three Gaussian peaks. 

\begin{figure*}[t]
  \centering
  \begin{subfigure}{0.38\textwidth}
  \begin{quantikz}
      & \lstick{$\ket{0}$}  & \gate[5]{\textbf{I}} & \gate[5]{\textbf{II}}  & \ \ldots\ \qw & \gate[5]{\textbf{II}} & \qw\\
      & \lstick{$\ket{0}$}  & & & \ \ldots\ \qw & & \qw\\
      & \lstick{$\ket{0}$}  & & & \ \ldots\ \qw & & \qw\\
      & \lstick{$\ket{0}$}  & & & \ \ldots\ \qw & & \qw\\
      & \lstick{$\ket{0}$}  & & & \ \ldots\ \qw & & \qw
  \end{quantikz}
  \label{subfig: dqs whole circuit}
  \end{subfigure}%
    \begin{subfigure}{0.62\textwidth}
        \begin{quantikz}
          &\gate[5]{\textbf{II}}&\qw\\
          &&\qw\\
          &&\qw\\
          &&\qw\\
          &&\qw
      \end{quantikz}\hspace{0.1cm}
    =
      \begin{quantikz}
          &\gate[5]{e^{-\iu V\Delta t}} & \gate[5]{\text{QFT}} & \gate[5]{e^{-\iu \frac{p^2\Delta t}{2}}} & \gate[5]{\text{iQFT}} & \qw\\
          & & & & & \qw\\
          & & & & & \qw\\
          & & & & & \qw\\
          & & & & & \qw
      \end{quantikz}
      \label{subfig: dqs iterations}
      \end{subfigure}
  \mcaption{Example of the quantum circuit (prior to measurements) for the DQS of the Schr\"{o}dinger equation in one dimension based on the method introduced by Zalka and Wiesner \cite{wiesner1996, zalka1998}. (Left) The whole quantum cicuit, composed of subcircuits \textbf{I} and \textbf{II}. Subcircuit \textbf{I} constructs the initial wave function of the simulation, and subcircuit \textbf{II} is applied $T/\Delta t$ times, where $T$ is the total duration of the simulation and $\Delta t$ is a sufficiently small discretized time step chosen such that $T/\Delta t$ is a positive integer. (Right) Subcircuit \textbf{II} begins by applying the diagonal unitary operator generated by the potential term of the Hamiltonian, followed by a QFT. The unitary operator generated by the kinetic term is then also diagonal, expressed in terms of the momentum operator $p$. Applying the diagonal unitary generated by the kinetic term and the inverse QFT then completes the iteration.}
  \label{fig: qc with Z-W}
\end{figure*}

\begin{figure}[t]
  \begin{subfigure}[t]{0.9\columnwidth}
  \centering
      \begin{quantikz}
          \lstick[wires=5]{$n $}\hspace{2mm}
          & &\gate[5]{e^{-\iu H_f}}\gategroup[2,steps=1,style={dashed,
rounded corners,fill=blue!40, inner xsep=2pt},
background]{} & \qw\\
          & & & \qw\\
          & \lstick[wires=3]{$l $}\hspace{1mm}
          & \gategroup[3,steps=1,style={dashed,
rounded corners,fill=gray!40, inner xsep=2pt},
background,label style={label position=below,anchor=
north,yshift=-0.2cm}]{}& \qw\\
          & & & \qw\\
          & & & \qw
      \end{quantikz}
      \end{subfigure}
  \begin{subfigure}[t]{\columnwidth}
  \centering
      \begin{quantikz}
        \lstick[wires=5]{$n $}\hspace{2mm}
          & & \qw& \gate[7]{e^{-\iu \widetilde H_f}}\gategroup[5,steps=1,style={dashed,
rounded corners,fill=blue!40, inner xsep=2pt},
background]{} & \qw& \qw&\\
          & & \qw & \qw & \qw & \qw&\\
          & \lstick[wires=3]{$l $}\hspace{1mm} & \gate[5]{U_S}& & \gate[5]{U_S^{\dagger}} & \qw&\\
          & & & & & \qw&\\
          & & & & & \qw&\mathrm{\hspace{0.35cm}}\\
          \lstick[wires=2]{$m $}\hspace{1mm} &
          \lstick{$\ket{0}$} & & \gategroup[2,steps=1,style={dashed,
rounded corners,fill=gray!40, inner xsep=2pt},
background,label style={label position=below,anchor=
north,yshift=-0.2cm}]{} & &\qw &\lstick{$\ket{0}$}\\
          &\lstick{$\ket{0}$} & & &  & \qw &\lstick{$\ket{0}$}
      \end{quantikz}
      \end{subfigure}
  \mcaption{General structures of quantum circuits that implement piecewise polynomial functions with the (top) ancilla-free and (bottom) ancilla-assisted approaches, using an example with $n = 5$, $l = 3$, and $m = 2$. In both cases, the qubits colored in gray are used to distinguish the subdomains and those colored in blue are used to implement the corresponding polynomial functions. (Top) The ancilla-free approach is based on labeling the subdomains by the $l$ most significant bits of the quantum registers that encode $x_k$. (Bottom) When only $m < l$ bits are needed to label the subdomains, we can apply a labeling operation $U_S$ to map the labels in the $l$ most significant bits of the main register into an $m$-qubit anillary label register, and uncompute it with $U_S^{\dagger}$ after using it to implement the piecewise function.}
  \label{fig: qc for both methods}
\end{figure}

\section{Theory}\label{sec:theory}
\subsection{Notation and definitions}\label{section : notations and preliminaries}
Consider an $n$-qubit quantum register with associated Hilbert space $\mathcal{H}_{n}$ of dimension $N := 2^n$. Without loss of generality, suppose that the one-dimensional spatial coordinate is restricted to the unit interval\footnote{We consider the generalization to multiple dimensions in a later section} $x \in \Omega = [0,1)$. In addition, define $\mathcal{F}_\alpha \subseteq \{ \Omega \longrightarrow \mathbb{R} \}$ as the vector space consisting of degree-$\alpha$ polynomials. Discretizing $\Omega$ using a grid of $N$ uniformly separated mesh points $x_k = k/N$ for $k \in \{0,1,\ldots, N-1\}$, we can encode wave function $\psi(t,x_k)$ as the amplitude of the computational basis element $|k\rangle$, where $k$ can be expanded in binary form as $k \equiv \sum_{a=1}^n k_{a}2^{n-a}$ for $k_{a} \in \{0,1\}$.

It will prove useful to introduce the following notation. Define $1 \leq L \leq N$ uniform cells in $\Omega$ as
\begin{equation}
    B_r := \Big[\frac{r}{L}, \frac{r+1}{L}\Big),
    \label{def. uni subdomain}
\end{equation}
where the index $r \in \Gamma := \{0,1,\ldots, L-1\}$ and $L = 2^l$ for some non-negative integer $l$. It is important to note that the integer given by the first $l$ bits in the binary representation of $k$ corresponds to the value of $r$ such that $x_k \in B_r$. Moreover, we consider a partitioning of the set $\Gamma$ into $1 \leq K \leq L$ blocks $\{\Gamma^{(s)}\}_{s=0}^{K-1}$ such that $\bigsqcup_{s=0}^{K-1}\Gamma^{(s)} = \Gamma$. We then introduce an associated partition of $\Omega$ as
\begin{equation}
    A_s := \bigsqcup_{r \in \Gamma^{(s)}} B_r,
    \label{def. ada subdomain}
\end{equation}
which define subdomains with the property that
\begin{equation}
    \bigsqcup_{s=0}^{K-1}A_s = \bigsqcup_{r=0}^{L-1}B_r = \Omega.
\end{equation}

\subsection{Classical preliminaries}\label{section : classical preliminaries}
We will consider piecewise polynomial functions in this article, which we classify as either uniform or adaptive, depending on the position of their knots.

Consider a function $f : \Omega \longrightarrow \mathbb{R}$, which is defined piecewise in terms of $1 \leq K \leq N$ constituent polynomial sub-functions $\{f_s \}_{s=0}^{K-1} \subseteq \mathcal{F}_\alpha$ as
\begin{equation}\label{e:piecewise}
    f(x) = \sum_{s=0}^{K-1}f_s(x)\mathbbm{1}_{A_s}(x),
\end{equation} 
where $\mathbbm{1}_{A_s} : \Omega \longrightarrow \{0,1\}$ is the indicator function for the subdomain $A_s \subseteq \Omega$. Unlike the superposition function $\sum_{s=0}^{K-1}f_s \in \mathcal{F}_\alpha$, we emphasize that $f$ need not be a member of $\mathcal{F}_\alpha$. In the special case where the number $K$ of subdomains equals the number $L$ of uniform cells, we refer \eqref{e:piecewise} as a \emph{piecwise uniform function}. Otherwise we refer to \eqref{e:piecewise} as a \emph{piecewise adaptive function}. 

Moreover, define for convenience, a function $S: \{0,1\}^n \rightarrow \{0,1,\ldots, K-1\}$ which returns the index of the subdomain containing the mesh point $x_k \in A_{S(k)}$,
\begin{equation}
    S(k) := \sum_{s=0}^{K-1}s\mathbbm{1}_{A_s}(x_k).
    \label{eq: labeling func}
\end{equation}
The key to our quantum algorithm is the following (proved in \Cref{app. gen eig eq proof}):
\begin{lemma}\label{lem:general}
\textit{ Let $m := \lceil\log_2{(K)}\rceil$ be the number of bits required to represent $K$ in binary, and $M:=2^m$. Then for all $k \in \{0,1\}^n$,}
\begin{subequations}\label{eq. general formula}
\begin{equation}
     f(x_k) = \sum^{M-1}_{t=0} g_{t}(x_k)\prod_{a=1}^{m}(-1)^{S_a(k)t_a},
     \label{eq. general formula 1}
\end{equation}
\textit{where $t_a$ denotes the $a$th bit in the binary expansion of the integer $t = \sum_{a=1}^{m}t_a2^{m-a}$ and $g_t : \Omega \longrightarrow \mathbb{R}$ is the function defined by}
\begin{equation}
     g_{t}(x) := \frac{1}{M} \sum^{K-1}_{s=0}(-1)^{s\cdot t}f_s(x),
     \label{eq. general formula 2}
\end{equation}
\textit{where $s\cdot t := \sum^{m}_{a=1}s_at_a$.}
\end{subequations}
\end{lemma}
Although \Cref{lem:general} does not require that the sub-functions $\{f_s\}$ are degree-$\alpha$ polynomials, in the special case where $f_s \in \mathcal{F}_\alpha$ $\forall s$, it follows immediately that $g_t \in \mathcal{F}_\alpha$ by closure of $\mathcal{F}_\alpha$ under linear combination. \Cref{lem:general} is evidently trivial when $K=1$, since then $f \in \mathcal{F}_\alpha$ is itself a degree-$\alpha$ polynomial. In the opposite limit where every mesh point $x_k$ is contained in a distinct subset $A_s$, the piecewise structure of $f$ is not resolved by the grid and \Cref{lem:general} then coincides with the Boolean Fourier inversion formula \cite{hadfield2021}.

\subsection{Problem statement}\label{section : quantum implementation}
Consider the problem of implementing the diagonal unitary operator $e^{-\iu H_f}$ on $\mathcal{H}_n$, where $H_f$ is defined by
\begin{equation}
    H_f |k\rangle = f(x_k)|k\rangle,
\end{equation}
for all $k$. In this work, we consider both ancilla-free and ancilla-assisted implementations, where the latter utilizes a workspace of $m$ additional qubits to reduce the circuit depth (shown in \Cref{fig: qc for both methods}). In the ancilla-free approach, we directly utilize \Cref{lem:general} to obtain a representation of $H_f$ suitable for an efficient quantum circuit construction. In the ancilla-assisted approach, an extra step involves a unitary $U_{S}$ acting on $\mathcal{H}_{n+m}$, which implements the labeling function in \eqref{eq: labeling func} for all $(k,u) \in \{0,1\}^{n+m}$,
\begin{equation}
    U_{S} \, |k \rangle |u\rangle = |k \rangle |S(k) \oplus u\rangle.
\end{equation}
We then define a Hamiltonian operator $\widetilde H_f$  on $\mathcal{H}_{n+m}$, again for all $(k,u) \in \{0,1\}^{n+m}$ as
\begin{equation}
\widetilde H_f | k \rangle |u\rangle = \sum_{t=0}^{M-1} g_t(x_k) \prod_{a=1}^{m} (-1)^{u_at_a} | k \rangle |u\rangle. 
\end{equation}
Using \Cref{lem:general}, the restriction of $\widetilde H_f$ to the subspace of states spanned by $|k\rangle |S(k)\rangle$ for all $k$ is given by
\begin{equation}\label{e:h_tilde_f}
\widetilde H_f | k \rangle |S(k)\rangle = f(x_k) \, | k \rangle |S(k)\rangle.
\end{equation}
Thus, by initializing the ancillary register in the state $|0\rangle^{\otimes m}$, we find that the required operators are given for all $|\psi\rangle \in \mathcal{H}_n$ by
\begin{align}
    (e^{-\iu H_f} |\psi\rangle) |0\rangle^{\otimes m}
    & = U_S^{\dagger} e^{-\iu \widetilde H_f} U_{S} \, |\psi\rangle |0\rangle^{\otimes m},
\end{align}
where $U_S^{\dagger}$ uncomputes the labels encoded by $U_S$. Both the ancilla-free and ancilla-assisted algorithms that we propose can be applied to piecewise polynomial functions with either uniform or adaptively chosen subdomains. In the case of adaptive grids, we will find that a compromise is struck between the utility of these methods depending on the details of the potential. No such compromise exists for uniform subdomains, however, and the ancilla-free approach is always superior. The detailed justification of these claims in the case of degree-2 polynomials is deferred to \Cref{app. qc complexity}.

\subsection{Relationship with previous work}\label{sec:previous}
In the limit where the number of subdomains equals the grid size ($K=N$), the Boolean Fourier inversion formula provides an exact yet inefficient ancilla-free implementation of $e^{-\iu H_f}$ \cite{hadfield2021}. In our work, we improve upon previous results by considering piecewise degree-$\alpha$ polynomial approximations. The effective piecewise constant approximation performed in \cite{welch2014}, which arises when resampling the underlying (approximated) function at rates lower than $N$, is generally a different approach from choosing $\alpha=0$ in the methods proposed in this article. While our method can also be applied with $\alpha = 0$, the regime where $1< K \ll N$ can be reached when $\alpha > 0$, allowing subdomains to be distinguished with fewer quantum gates. A method to perform basis embedding of a piecewise smooth function into a quantum register has been introduced in \cite{haner2018}, which uses an ancillary label register to distinguish the subdomains. By combining this with the method of phase kickback, one can perform DQS with the use of $(\alpha+1)n + m + 1$ total qubits if no non-trivial pebbling strategy \cite{bennett1989, parent2015} is applied. Our ancilla-assisted method, in contrast, requires only $n + m$ total qubits, and we offer additional flexibility in the form of an ancilla-free implementation.

In \Cref{section : num exp}, we will apply our method to the cosine potential (among other examples). An exact implementation of the time evolution operator for this potential given in \cite{georgeot2001, levi2004} requires multiple ancillary registers with at least $n$ qubits each, and involves a total gate count of $O(\log{(N)}^3)$. Two approximate implementations were also given in \cite{levi2004}, where the ``Chebyshev polynomial'' method and the ``slice method'' are both economical in terms of the number of ancilla qubits. The former involves a total of $O(\log{(N)}^{\alpha_{\text{c}}})$ operations, where $\alpha_{\text{c}}$ is the degree of the Chebyshev polynomial approximation. When fixing the precision of approximation, the latter can also be efficiently computed, where the asymptotic gate count of the whole iteration is dominated by the QFT. This is also true for our method. Note that while our method and the Chebyshev polynomial method are applicable to a universal class of potential functions in real-space, the slice method utilizes properties of the cosine function which generally are not satisfied by other functions. Furthermore, no quantum gates are needed to distinguish subdomains if approximating the potential function with Chebyshev polynomials, but when the precision of approximation is fixed, $\alpha_{\text{c}} \geq \alpha$, so our method could involve fewer quantum gates for complicated potential functions.

\section{Quantum circuit construction}\label{sec:examples}

As explained in the introduction, we aim to evolve the state vector according to the time-dependent Schr\"{o}dinger equation in which the potential function is replaced by a piecewise polynomial approximator. Suppose that we are given a classical algorithm that performs such approximation for some chosen error tolerance, returning either a uniform or adaptive piecewise polynomial function. This section is concerned with implementing the diagonal unitary operator generated by the piecewise polynomial function on a quantum computer. To this end, we demonstrate the explicit construction of quantum circuits for both the ancilla-free and ancilla-assisted approaches, followed by a discussion of their extensions to higher dimensions.

\subsection{Ancilla-free implementation}\label{subsection : algorithm for 1D}
Consider the partition of the fundamental domain $\Omega = [0,1)$ into uniform subdomains, such that the subdomains correspond to the uniform cells, and thus $K = L = M$. In this case, the target unitary can be implemented without introducing an ancillary register by using the fact that the binary representation of $S(k)$ can be represented by the first $m=l$ bits of $k \in \{0,1\}^n$. That is,
\begin{equation}\label{e: uniform label}
    S_a(k) = k_a \quad \quad a=1,\ldots, m,
\end{equation}
so that $(-1)^{S_a(k)} \in \{-1,1\}$ is the eigenvalue of the Pauli operator $Z_a$ corresponding to the eigenvector $\ket{k_a}$. By replacing $(-1)^{S_a(k)}$ by $Z_a$ and re-indexing summations, \eqref{eq. general formula 1} can be further simplified to the resulting Hamiltonian
\begin{equation}\label{e:efficient}
     H_f = \sum^{M-1}_{t=0} H^{(t)}\prod_{a=1}^{m}Z^{t_a}_{a}.
\end{equation}
Here, $H^{(t)}$ is an $\alpha$-local Hamiltonian constructed from $Z_{m+1},\ldots,Z_n$, which, in the specific case of $\alpha = 2$, takes the explicit form (derived in \Cref{app. gen uniform ham rep proof})
\begin{equation}\label{e:h_f}
     H^{(t)} = \frac{1}{2}\bigg{[}\sum_{b=m+1}^{n}\bigg{(}\sum_{c=b+1}^{n}\theta^{(t)}_{bc}Z_c + \phi^{(t)}_{b}\bigg{)}Z_b + \lambda^{(t)}\bigg{]},
\end{equation}
where $\theta^{(t)}_{bc},\phi^{(t)}_{b}, \lambda^{(t)} \in \mathbb{R}$ can be classically precomputed, and the factor of $1/2$ is conventional. The diagonal unitary $e^{-\iu H_f}$ is thus
\begin{align}\label{e:separated U}
    e^{-\iu H_f} &= \prod_{t=0}^{M-1} U^{(t)}\\
    &= \prod_{t=0}^{M-1}\bigg{[}\prod_{b=m+1}^{n}\bigg{(}\prod_{c=b+1}^{n} U^{(t)}_{b,c}(\theta)\bigg{)} U^{(t)}_{b}(\phi) \bigg{]} U^{(t)}(\lambda),\nonumber
\end{align}
where the unitary operators $U^{(t)}_{b,c}(\theta)$, $U^{(t)}_{b}(\phi)$, and $U^{(t)}(\lambda)$ each depend on the angles $\theta$, $\phi$, and $\lambda$ in \eqref{e:h_f}. Each of these unitary operators can be implemented based on so-called ``CNOT staircases'' \cite{whitfield2011, yordanov2020, yordanov2021} (see \Cref{app. cnot staircase}), and circuit depth can be considerably reduced by maximizing the number of neighboring CNOT gates, which can be cancelled in pairs. Since all terms in \eqref{e:efficient} commute with each other, the corresponding unitary operators in \eqref{e:separated U} can be applied in an arbitrary order. This implies that we can first group the unitaries with the same value of $t$ together, so that all internal CNOT staircases cancel amongst themselves. For a given value of $t$, we similarly group terms with the same value of $b$ in order to exploit further cancellations. Finally, CNOT gates can also be cancelled between $U^{(t_1)}$ and $U^{(t_2)}$ for $t_1 \neq t_2$. For this last step, it was shown in \cite{welch2014} that the order in which the maximal number of CNOT gates can be cancelled is the sequency order, given by the Gray code \cite{beauchamp1985}. Optimizing the quantum circuit as described above (in particular, with $\alpha = 2$) leads to $O(M\log{(N/M)}^2)$ rotation-Z and CNOT gates (see \Cref{app. qc complexity 1} for details).

The fact that the quantum circuit is built entirely based on CNOT staircases motivates one to save circuit depth further by ignoring small Pauli-Z rotations, since we can then also remove all the CNOT gates associated with them. To this end, we introduce an adjustable threshold $\tau \geq 0$ such that we ignore all rotations with angles $\mu$ that satisfy $|\mu| < \tau$ (no gate is removed when $\tau = 0$).

We remark that the method can also be applied to piecewise polynomials with non-uniform subdomains by viewing the cells in \eqref{def. uni subdomain} as the subdomains instead of their union in \eqref{def. ada subdomain}, and neglecting the fact that some polynomial functions are equal in different cells. The gate counts associated with this scenario are descibed in \Cref{app. qc complexity 4}.

\subsection{Ancilla-assisted implementation}\label{subsection : piecewise adaptive algorithm}

Consider now a decomposition of $\Omega = [0,1)$ into non-uniform subdomains $A_s$, each comprised of the elemental uniform cells $B_r$ $\forall r \in \Gamma^{(s)}$, as described in \Cref{section : notations and preliminaries}. We apply the desired diagonal unitary by introducing an ancillary label register with $m$ qubits indexed from $n+1$ to $n+m$, and refer to the quantum register composed of qubits $1$ to $n$ as the main register. The labeling unitary, $U_{S}$, can be implemented by applying conditional gates which encode $S(k)$ to the target label register, conditioned on the value $r = \sum_{a=1}^l k_{a}2^{l-a}$ satisfying $r \in \Gamma^{(S(k))}$. Once the main and label registers are entangled in this way, $(-1)^{S(k)}$ in \eqref{eq. general formula 1} can now be translated into Pauli-Z operators acting on the label register as in the ancilla-free implementation, so that we obtain
\begin{equation}
     \widetilde H_f = \sum^{M-1}_{t=0} \widetilde H^{(t)}\prod_{a=1}^{m}Z^{t_a}_{n+a},
\end{equation}
after re-expressing \eqref{eq. general formula 2} in terms of Pauli-Z operators. Here, $\widetilde H^{(t)}$ is an $\alpha$-local Hamiltonian constructed from $Z_{1},\ldots,Z_n$, which, in the specific case of $\alpha = 2$ takes the explicit form (derived in \Cref{app. gen uniform ham rep proof})
\begin{equation}\label{eq. ham rep adaptive}
     \widetilde H^{(t)} = \frac{1}{2}\bigg{[}\sum_{b=1}^{n}\bigg{(}\sum_{c=b+1}^{n}\widetilde{\theta}^{(t)}_{bc}Z_c + \widetilde{\phi}^{(t)}_{b}\bigg{)}Z_b + \widetilde{\lambda}^{(t)}\bigg{]},
\end{equation}
where we again defined $\widetilde{\theta}^{(t)}_{bc} \widetilde{\phi}^{(t)}_{b}, \widetilde{\lambda}^{(t)} \in \mathbb{R}$, corresponding to angles of rotation gates. Unlike \eqref{e:h_f}, the summations now include the full range $1,\ldots,n$. This implies that the ancilla-assisted method introduces additional gates compared to the ancilla-free method (and an additional $m$ ancilla qubits) for fixed $M$. However, an adaptive piecewise polynomial approximation can generally lead to much smaller $M$ than a uniform approximation for a fixed precision of approximation, resulting in a reduced circuit depth. This point will be revisited in \Cref{app. qc complexity 4}, where explicit gate counts are analyzed, and within a specific example in \Cref{subsubsection : num exp results eckart}.

While not necessarily an optimal implementation, the labeling unitary $U_{S}$ can be implemented with $O(L\log{(M)})$ single- and two-qubit gates, and we use this implementation whenever explicit gate counts are discussed in this article (see \Cref{app. labeling gates}). The target unitary $e^{-\iu \widetilde{H}_f}$ can then be applied in a similar fashion as in the ancilla-free implementation, that is, by ordering the individual unitary operators in a way that cancels as many CNOT gates as possible. In the final step, the labeling is uncomputed, so that we disentangle the main register from the label register, and reuse the label register for labeling subdomains in subsequent iterations in the DQS algorithm. This part requires the same number of gates as the labeling part. Excluding the labeling operation and its uncomputation, the ancilla-assisted implementation utilizes $O(M\log{(N)}^2)$ rotation-Z and CNOT gates (see \Cref{app. qc complexity 2} for details).

\subsection{Extension to multiple dimensions}\label{subsection : higher D}

The discussion so far focused on the one-dimensional Schr\"{o}dinger equation. The method introduced in this work can be generalized to $d$ dimensions simply by considering the coordinate $x$ in \Cref{section : classical preliminaries} as a tuple containing all $d$ coordinates $x = (x^{(1)}, x^{(2)}, \ldots, x^{(d)})$. \Cref{lem:general} still holds, where the grid $x_k$ is to be understood as a tuple that holds each discretized coordinate $k^{(i)}/N^{(i)}$ for $k^{(i)}\in \{0,1,\ldots, N^{(i)}-1\}$. Here, we have introduced $d$ quantum registers each encoding their corresponding discretized coordinate with $n^{(i)}$ qubits such that $n = \sum_{i = 1}^d n^{(i)}$, along with their associated Hilbert space dimensions $N^{(i)} := 2^{n^{(i)}}$.

As before, we consider uniform cells $B_r^{(i)}$ in each coordinate, defined as in \eqref{def. uni subdomain} with possibly distinct $L^{(i)}$. We then consider a multivariate piecewise polynomial function defined on cartesian products of $B_r^{(i)}$ $\forall i \in \{1,2,\ldots, d\}$. The idea is the same as in the one-dimensional case, but now we group together $d$-dimensional cells instead of uniform lines to form $K$ subdomains. Thus, we will have $d$ main registers each storing a coordinate, and for the ancilla-assisted method, an additional label register with $m = \lceil \log_2{(K)}\rceil$ qubits. The quantum gates would then be applied in a similar fashion as in the one-dimensional case, but now with entangling gates performed across these quantum registers.

In the ancilla-free implementation, \Cref{lem:general} can be restated in an expression that clarifies the construction of the quantum circuit (as shown in \Cref{app. lemma for uniform}), the result of which has $d$ main registers each storing a coordinate in the same way as in the 1-dimension case.

Embedding this method into the Zalka-Wiesner algorithm can be done straightforwardly by performing the QFT (and its inverse) separately onto each quantum register; thereby enabling the DQS in real space for arbitrary potential functions.

\section{Numerical experiments}\label{section : num exp}
The quantum circuits introduced in \Cref{subsection : algorithm for 1D} and \Cref{subsection : piecewise adaptive algorithm} can be applied to approximate an arbitrary one-dimensional potential function $V$. In this section, we will use three example potential functions to illustrate the methods, fixing the degree to $\alpha = 2$. We distinguish between two types of numerical experiments. The first is a construction that allows for the analysis of gate counts and errors when $\tau > 0$, and the second is to perform DQS with the Zalka-Wiesner algorithm. In both cases, we use an ideal (or noiseless) classical simulator of a quantum computer \cite{aleksandrowicz2019qiskit}.

\subsection{Setting} \label{subsection : setting}

\subsubsection{Numerical experiment 1} \label{subsubsection : NE1}
In the first type of experiment, we will only consider the ancilla-free implementation. Consider an input potential function $V:\Omega\longrightarrow\mathbb{R}$ and a piecewise quadratic $\epsilon$-approximator $f^{\epsilon}:\Omega\longrightarrow\mathbb{R}$ (we make the dependence on $\epsilon$ explicit in the notation here) of $\tilde{V} := V \Delta t$ with respect to the infinity norm,
\begin{equation}
    \Vert \tilde{V} - f^{\epsilon} \Vert_\infty \leq \epsilon.
\end{equation}
Let $U(f^{\epsilon},\tau)$ denote the quantum circuit used in the construction of the unitary operator $e^{-\iu H_{f^{\epsilon}}}$, where $\tau$ denotes the threshold parameter. In this section, we calculate the approximation errors
\begin{align}
\begin{split}
    \delta (\tau) & := \big\Vert U(f^{\epsilon},\tau) - e^{-\iu \tilde{V}}\big\Vert \\
    \widetilde\delta(\tau) & := \big\Vert U(f^{\epsilon},\tau) - e^{-\iu H_{f^{\epsilon}}}\big\Vert,
\end{split}
\end{align}
where the spectral norm is applied. Since $U(f^{\epsilon},\tau)$, $e^{-\iu \tilde{V}}$, and $e^{-\iu H_{f^{\epsilon}}}$ are all diagonal operators, the two errors can be obtained by
\begin{align}
\begin{split}
    \delta (\tau) & = 2^{n/2}\big\Vert \ket{\psi_{\epsilon, \tau}^{\text{qc}}} - e^{-\iu \tilde{V}} | + \rangle^{\otimes n}
    \big\Vert_\infty \\
    \widetilde\delta(\tau) & = 2^{n/2}\big\Vert \ket{\psi_{\epsilon, \tau}^{\text{qc}}} - e^{-\iu H_{f^{\epsilon}}} | + \rangle^{\otimes n} \big\Vert_\infty,
\end{split}
\end{align}
where $\ket{+} = \frac{1}{\sqrt{2}}(\ket{0}+\ket{1})$ is the positive eigenstate of the Pauli-X operator and $\ket{\psi_{\epsilon, \tau}^{\text{qc}}} := U(f^{\epsilon},\tau) |+\rangle^{\otimes n}$.

Thus, after classically obtaining $f^{\epsilon}$, we prepare an equal superposition of computational basis states by applying a Hadamard gate onto each qubit, and then apply $U(f^{\epsilon},\tau)$ to obtain $\ket{\psi_{\epsilon, \tau}^{\text{qc}}}$ (as shown in \Cref{fig: qc for example}). For the purpose of analysis, we obtain the $N$-dimensional state vector instead of performing measurements as in a realistic quantum computing experiment. By fixing $\epsilon$ while varying $\tau$, this numerical experiment enables the analysis of the relationship between the removal of small rotations and the errors that arise from it.

\def\myvdots{\ \vdots\ }
\begin{figure}[t]
  \centering
  \begin{quantikz}
    \lstick[wires=4]{$n $}\hspace{2mm}
      & \lstick{$\ket{0}$}  & \gate{H}\slice{\hspace{0.5cm}$\ket{+}^{\otimes n}$} & \gate[4, nwires=3]{U(f^{\epsilon},\tau)}\slice{$\ket{\psi_{\epsilon, \tau}^{\text{qc}}}$} & \qw\\
      & \lstick{$\ket{0}$}  & \gate{H} & & \qw\\
      & \lstick{\myvdots} & \myvdots & & \myvdots &\\
      & \lstick{$\ket{0}$}  & \gate{H} & & \qw
  \end{quantikz}
  \mcaption{Quantum circuit used in Numerical experiment 1, ran on a classical simulator of a quantum computer. $n$ qubits are prepared as an equal superposition of all computational basis states by applying a Hadamard gate onto each qubit initialized to $\ket{0}$. This is followed by the quantum gates corresponding to the ancilla-free implementation of $e^{-\iu H_{f^{\epsilon}}}$, denoted as $U(f^{\epsilon},\tau)$, which produces the final state $\ket{\psi_{\epsilon, \tau}^{\text{qc}}}$.}
  \label{fig: qc for example}
\end{figure}

\subsubsection{Numerical experiment 2} \label{subsubsection : NE2}

The second type of numerical experiment that we perform is to apply $U(f^{\epsilon},\tau)$ as a sub-routine in the Zalka-Wiesner algorithm, where we now consider both the ancilla-free and ancilla-assisted implementations. The quantum circuit would then be the one shown in \Cref{fig: qc with Z-W}, where we are given the initial state and use our approximate methods to implement the potential term. The kinetic term corresponds to a diagonal unitary generated by a quadratic polynomial function, so it can be implemented straightforwardly. We again obtain the $N$-dimensional state vector instead of measured data, but now we store the state vectors at each iteration to observe the dynamics of a quantum particle in the potential.

For the two examples on which we perform this numerical experiment, we consider a grid such that $n  = 10$, and we discretize the evolution time up to $T = 0.6$ into $100$ time steps such that $\Delta t = 0.006$. A quantum particle is being initialized as a normalized Gaussian wave packet such that $\psi(x,0) \propto e^{-(x-\bar{x})^2/2\bar{\sigma}^2 + i\bar{p}(x-\bar{x})}$ for $(\bar{x}, \bar{p}, \bar{\sigma}) = (-3, 10, 0.5)$. Since we are not interested in analyzing the role of a finite $\tau$ in this numerical experiment, we fix $\tau = 0$ to consider the full implementation.

\subsection{Results and discussion} \label{subsection : num exp results}
\subsubsection{Cosine} \label{subsubsection : num exp results cos}
First, we perform Numerical experiment 1 by approximating $\tilde{V}_{\text{cos}}(x) = \cos{(x)}$ as a piecewise quadratic function, where $x \in [-\pi, \pi)$. The cosine potential is relevant for the DQS of, e.g., the quantum kicked rotor model \cite{georgeot2001, levi2003}, the quantum kicked Harper model \cite{levi2004}, and the Kogut-Susskind model with $U(1)$ gauge group \cite{sohaib2022}. The goal of the numerical experiment is to understand the relationship between the approximation error and the amount of resources used in this example.

\begin{figure}[t]
  \centering
  \includegraphics[width=\columnwidth]{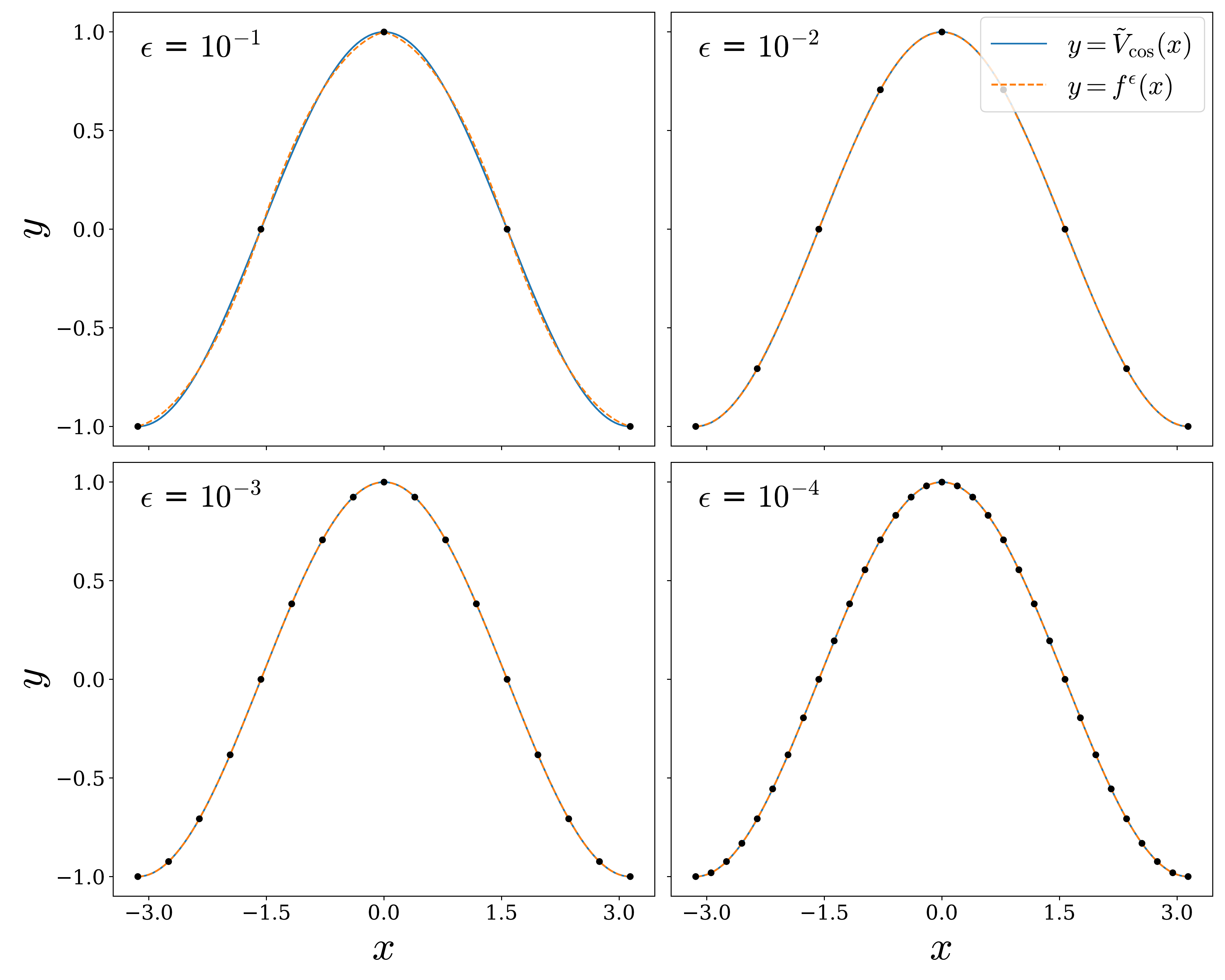}
  \mcaption{The piecewise quadratic $\epsilon$-approximators $f^\epsilon(x)$ obtained classically for $\epsilon \in \{10^{-1}, 10^{-2}, 10^{-3}, 10^{-4}\}$, along with the exact function $\tilde{V}_{\text{cos}}(x) = \cos{(x)}$. Due to symmetry in the cosine function, the resulting piecewise functions have uniform subdomains, indicated by the positions of the knots (black dots).}
  \label{fig: pw function and V for cos ex}
\end{figure}

\begin{figure}[t]
\centering
    \includegraphics[width=\columnwidth]{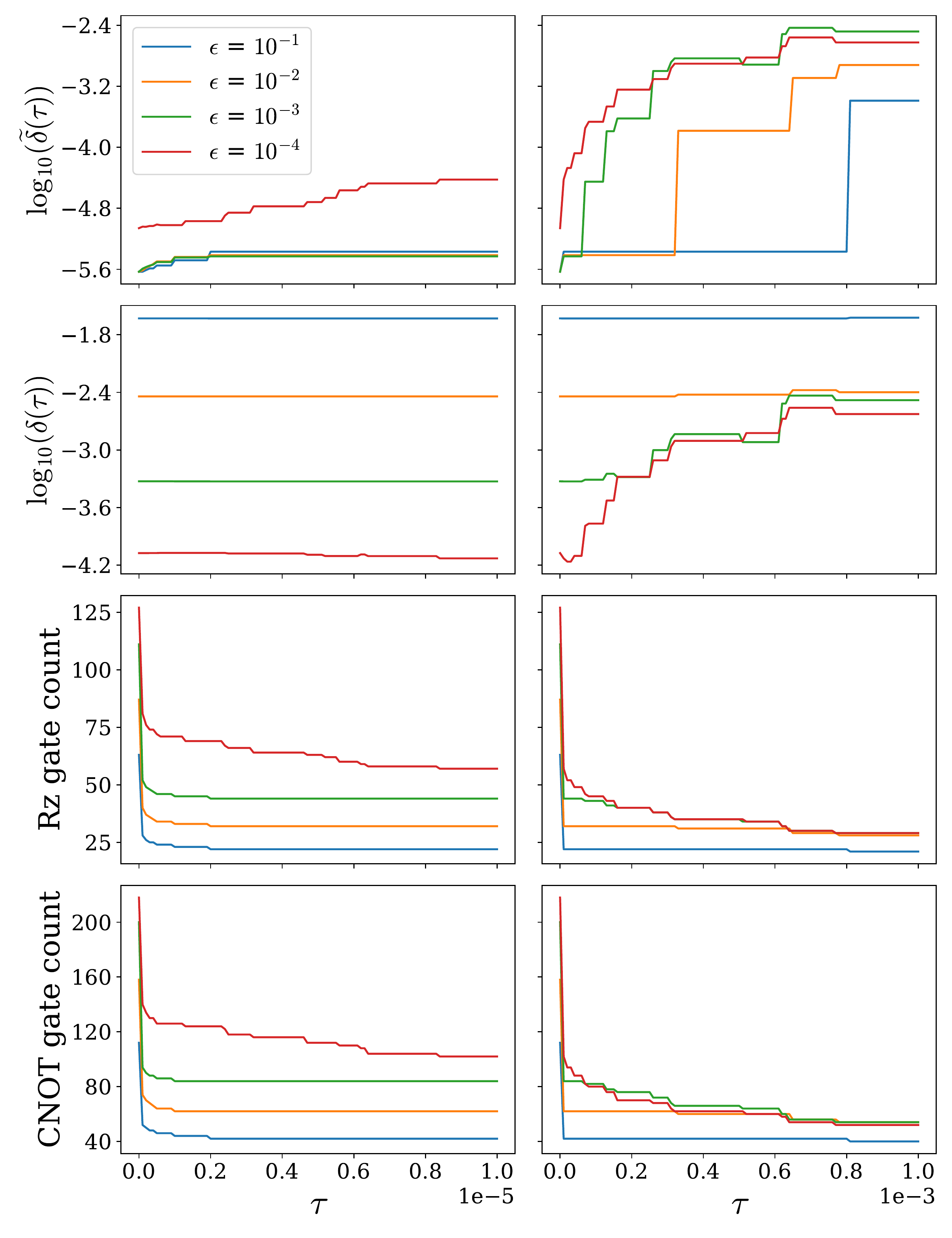}
  \mcaption{$\log_{10}{(\widetilde\delta(\tau))}$, $\log_{10}{(\delta(\tau))}$, rotation-Z gate count, and CNOT gate count as a function of (left column) $\tau \leq 10^{-5}$ and (right column) $\tau \leq 10^{-3}$. (Left column) Even though ignoring small rotations leads to a noticeable increase in $\widetilde\delta(\tau)$, the scale of which is negligible in $\delta(\tau)$, thus enabling the reduction of gates without affecting the approximation error with respect to $\tilde{V}$. (Right column) Increasing $\tau$ further generally leads to noticeable increases in $\delta(\tau)$, where the errors corresponding to different values of $\epsilon$ start to become comparable.}
  \label{fig: data for cos ex}
\end{figure}

We run the numerical experiment with $n = 7$. Fixing $\epsilon \in \{10^{-1}, 10^{-2}, 10^{-3}, 10^{-4}\}$, the four piecewise quadratic functions $f^{\epsilon}$ that we obtained (shown in \Cref{fig: pw function and V for cos ex}) give rise to $K = 4, 8, 16, 32$, respectively, and we vary $\tau \in [0, 10^{-3})$ for each of them (see \Cref{app. e.g. of num exp with m =2} for a detailed analysis of the case where $\epsilon = 10^{-1}$). The result of the numerical experiment is shown in \Cref{fig: data for cos ex}, where the right column shows the full result, and the left column magnifies the region $\tau \in [0, 10^{-5})$. Focusing first on the left column, we observe that for all $\epsilon$, while $\widetilde\delta(\tau)$ noticeably increases with $\tau$, the scale of which is small enough to have only a negligible impact on $\delta(\tau)$. Simultaneously, the gate counts for each $\epsilon$ also rapidly reduces for small $\tau$ until they saturate, allowing one to implement the desired operation with much fewer gates with only a negligible reduction in accuracy. On the right column, however, we can observe that the increase in $\delta(\tau)$ for $\epsilon = 10^{-2}, 10^{-3},$ and $10^{-4}$ becomes non-negligible as $\tau$ increases, becoming comparable between the three choices of $\epsilon$. Nevertheless, the circuit depth is still reduced further, allowing one to choose an optimal combination of $\epsilon$ and $\tau$ depending on the desired accuracy and available resources.

\subsubsection{Eckart barrier} \label{subsubsection : num exp results eckart}
\begin{figure*}[t]
  \centering
   \includegraphics[width=\textwidth]{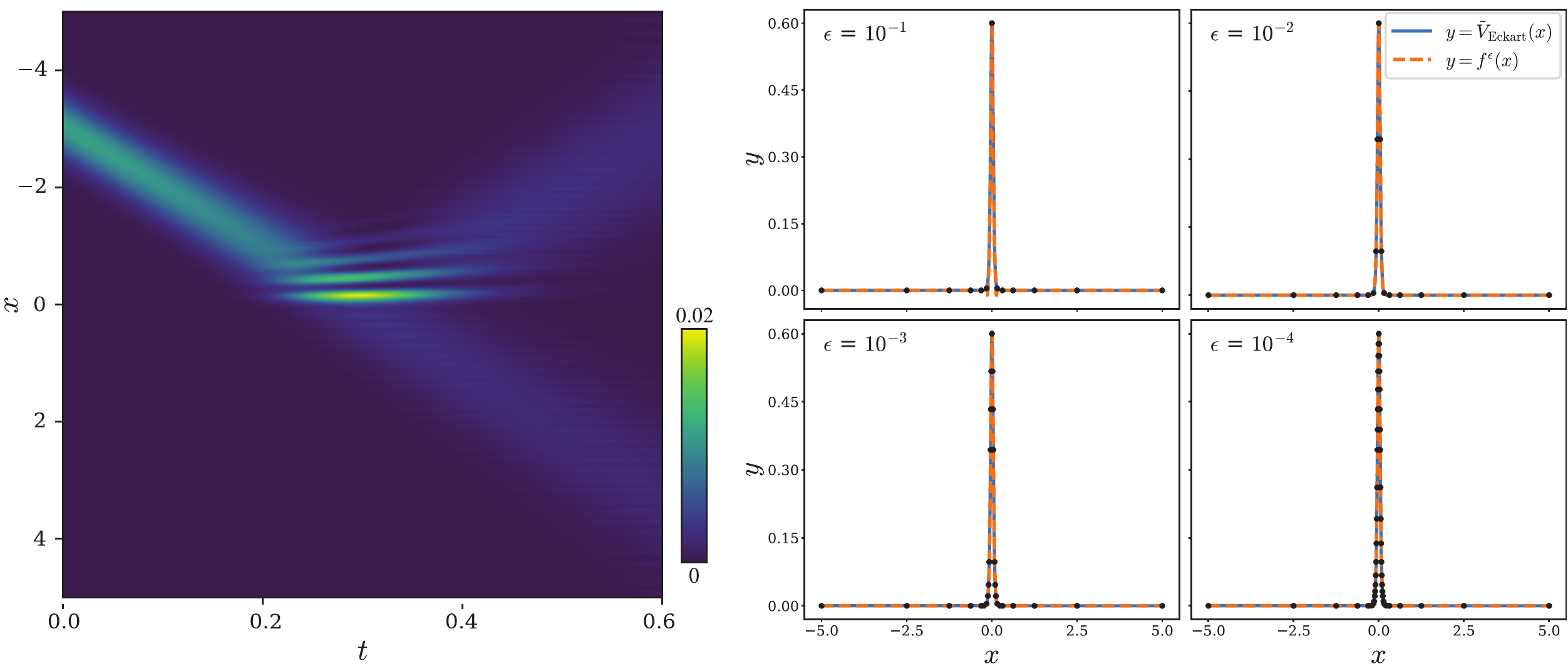}
   \mcaption{Potential function and the DQS result for the Eckart barrier example. (Left) $|\psi(x,t)|^2$ plotted as a result of the DQS of a quantum particle in the Eckart barrier potential, using an adaptive piecewise approximation with $\epsilon = 10^{-2}$. We can observe that, as expected, the wave function of the quantum particle separates into components due to reflection and transmission as it meets the potential barrier. (Right) The piecewise quadratic function $f^\epsilon(x)$ obtained classically for $\epsilon \in \{10^{-1}, 10^{-2}, 10^{-3}, 10^{-4}\}$ and the exact function $\tilde{V}_{\text{Eckart}}(x) = V_{\text{Eckart}}(x) \Delta t$. The resulting piecewise functions have non-uniform subdomains indicated by the positions of the knots (black dots).}
  \label{fig: eckart barrier example}
\end{figure*}

In the next example, we consider the Eckart barrier potential $V_{\text{Eckart}}(x) = \gamma\sech^2{(x/\nu)}$ for $x \in [-5, 5)$, where we choose $\gamma = 100$ and $\nu = 0.05$. A similar potential has been studied in \cite{welch2014}, where the authors used an effective piecewise constant approximation based on an expansion in the Walsh-Fourier basis to perform the DQS of a quantum particle. Here, we similarly simulate the evolution of a quantum particle under this potential (Numerical experiment 2), but we instead apply an adaptive piecewise quadratic approximation. The function $\tilde{V}_{\text{Eckart}}(x) := V_{\text{Eckart}}(x) \Delta t$ can be approximated to different orders of precision, and the resulting approximation functions are shown on the right hand side of \Cref{fig: eckart barrier example}. Note the adaptivity of the piecewise approximation, which could not be observed from \Cref{fig: pw function and V for cos ex} due to the symmetry in $\tilde{V}_{\text{cos}}(x)$.

First consider implementing the potential term with the ancilla-free method, where we choose $\epsilon = 10^{-2}$. With this choice, we obtain a piecewise quadratic function with $K = 16$, $M = 16$, and $L = 256$ (shown in the right hand side of \Cref{fig: eckart barrier example}). This function requires $l = \log_2{(L)} = 8$ qubits within the main register to label the cells, giving rise to a rotation-Z gate count of 4095 and a CNOT gate count of 7924. The result of performing the DQS with this approximation is given on the left hand side of \Cref{fig: eckart barrier example}, and shows features that are consistent with a classical simulation.

\begin{figure}[t]
  \centering
  \includegraphics[width=\columnwidth]{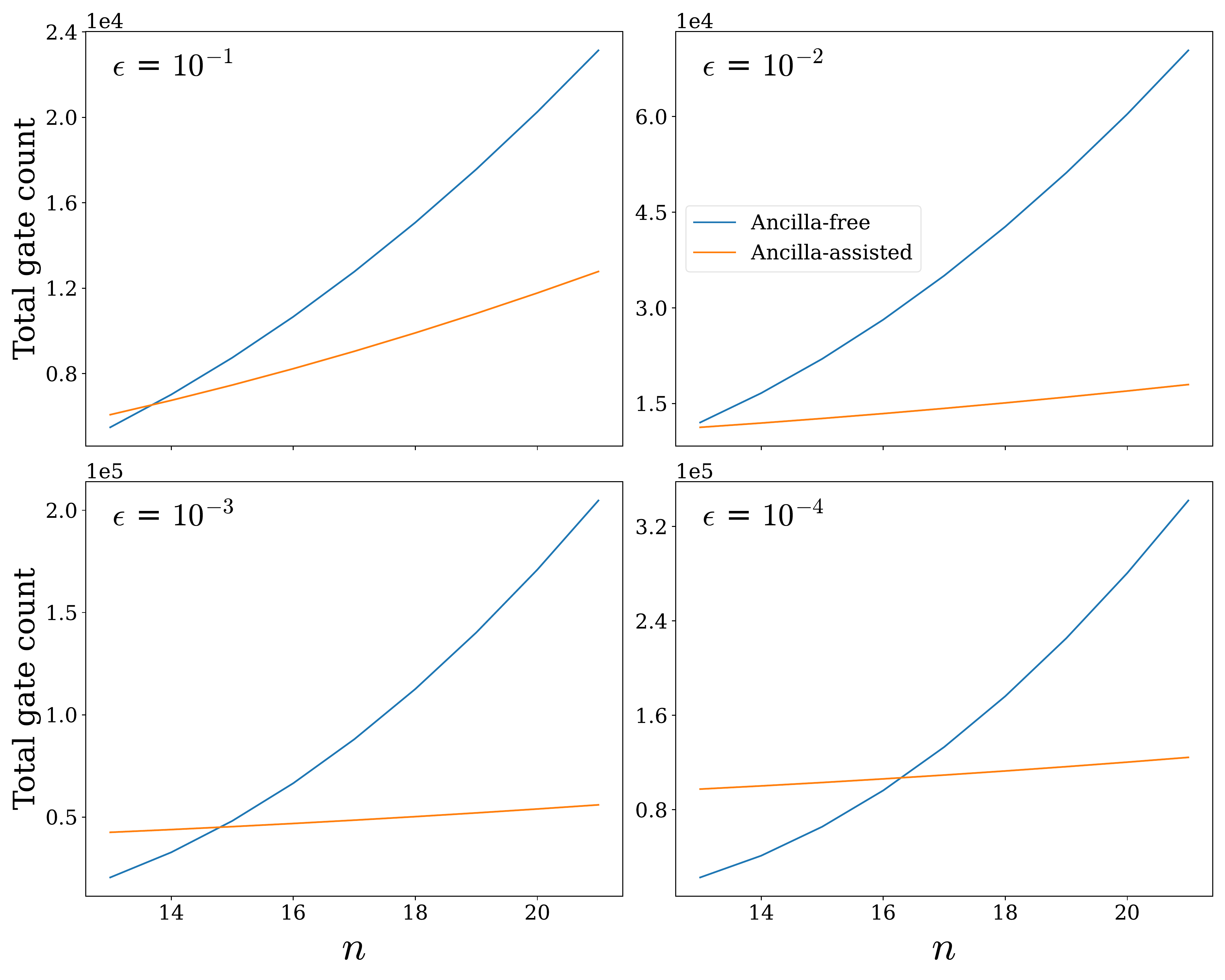}
  \mcaption{Scaling of gate count in Eckart barrier example for $\epsilon \in \{10^{-1}, 10^{-2}, 10^{-3}, 10^{-4}\}$, for an ancilla-free or ancilla-assisted implementation. The gate count for the ancilla-assisted implementation includes not only the polynomial evaluation but also the gates used for labeling and its uncomputation with the method given in \Cref{app. labeling gates}. For each precision of approximation, the total gate count of the ancilla-assisted implementation falls below that of the ancilla-free implementation for $n$ greater than $n^*(10^{-1}) = 13.70$, $n^*(10^{-2}) = 12.78$, $n^*(10^{-3}) = 14.82$, $n^*(10^{-4}) = 16.31$ up to two decimal places (see \Cref{app. qc complexity 4}).}
  \label{fig: gate count}
\end{figure}

The DQS can also be performed with the ancilla-assisted implementation, which accompanies different amount of resources. Instead of labeling within the main register, we introduce an ancillary label register with $m = \lceil\log_2{(K)}\rceil = 4$ qubits. For the same choice $\epsilon = 10^{-2}$, this method requires 895 rotation-Z gates and 1754 CNOT gates if we ignore the gates used for labeling and uncomputing the label, which is a considerable reduction compared to the ancilla-free implementation. If we include the labeling and its uncomputation performed using the quantum circuit presented in \Cref{app. labeling gates}, however, there will be 4080 controlled-Z gates, 2262 CNOT gates, 2048 rotation-X gates, and 1405 rotation-Z gates in total, involving more single- and two-qubit gates than the ancilla-free implementation. While the example considered here does not show the advantage of applying the ancilla-assisted implementation, the reduction in circuit depth compared to an ancilla free implementation can be observed when considering larger $n$. For fixed $\epsilon \in \{10^{-1}, 10^{-2}, 10^{-3}, 10^{-4}\}$,  \Cref{fig: gate count} shows the total gate counts (see \Cref{app. qc complexity} for details) associated with the ancilla-free and ancilla-assisted implementations of the approximated Eckart barrier potential. We can observe that the number of quantum gates in the ancilla-assisted method does not scale as fast as that in the ancilla-free method. This is due to the fact that $M$ is much smaller when viewing the piecewise function as defined on adaptive subdomains than on uniform subdomains, as illustrated in \Cref{fig. table of number of subdomains} for the specific case of $n = 20$. The difference is especially large in this example because of the sharp peak in the Eckart barrier potential, and the ancilla-assisted implementation is expected to considerably reduce the circuit depth of a general class of potential functions with highly non-uniform and distinctive features.

\begin{table*}[t]
\centering
\begin{tabular}{ccccc} \toprule
    {$\epsilon$} & {\begin{tabular}{c}M\\(ancilla-free)\end{tabular}} & {\begin{tabular}{c}M\\(ancilla-assisted)\end{tabular}} & {\begin{tabular}{c}Gate count\\(ancilla-free)\end{tabular}} & {\begin{tabular}{c}Gate count\\(ancilla-assisted)\end{tabular}} \\ \midrule
    $10^{-1}$  & $2^6$ & $2^4$ & $20257$ & $11776$\\
    $10^{-2}$  & $2^{8}$  & $2^4$ & $60389$ & $16960$\\
    $10^{-3}$  & $2^{10}$  & $2^5$ & $170985$ & $53948$\\
    $10^{-4}$  & $2^{11}$  & $2^6$ & $280555$ & $120248$\\
    \bottomrule
\end{tabular}
\mcaption{For the Eckart barrier potential in \Cref{subsubsection : num exp results eckart} with $\epsilon \in \{10^{-1}, 10^{-2}, 10^{-3}, 10^{-4}\}$ and fixed $n = 20$ : Values of $M$ when applying the ancilla-free method (in which case $M=L$) or the ancilla-assisted method, and their associated total gate counts (including the labeling operation and its uncomputation in the ancilla-assisted case with the method given in \Cref{app. labeling gates}). Intuitively, the advantage of the ancilla-assisted method arises due to the polynomial scaling of $K$ (which defines $M$) as opposed to the exponential scaling of $L$.}\label{fig. table of number of subdomains}
\end{table*}

\subsubsection{Multiple Gaussian peaks} \label{subsubsection : num exp results gaussians}
Finally, we consider an artificial potential function given by $V_{\text{Gauss}}(x) = \gamma_1\exp{\frac{-(x-\nu_1)^2}{2\kappa_1^2}} + \gamma_2\exp{\frac{-(x-\nu_2)^2}{2\kappa_2^2}} + \gamma_3\exp{\frac{-(x-\nu_3)^2}{2\kappa_3^2}}$, which is a sum of three Gaussian functions with parameters $(\gamma_1, \nu_1, \kappa_1) = (100, -2, 1/30)$, $(\gamma_2, \nu_2, \kappa_2) = (200/3, 0, 1/3)$, and $(\gamma_3, \nu_3, \kappa_3) = (25/3, 3, 1/2)$. The result of applying Numerical experiment 2 using an adaptive piecewise quadratic approximation is shown in \Cref{fig: triple gaussian example}, where we choose $\epsilon = 10^{-2}$. As expected, we observe both reflective and transmissive components of the wave function as the quantum particle meets the first two potential barriers.

\section{Conclusion}\label{section : conclusion}

Two methods to apply diagonal unitaries generated by piecewise polynomial functions on a quantum computer were introduced in this work. The first is an ancilla-free implementation suitable for uniform piecewise polynomials, while the other utilizes an ancillary quantum register, and can reduce the circuit depth for certain adaptive piecewise polynomials. Both algorithms can be applied in DQS, where piecewise polynomial approximations are performed on the potential function.

The methods introduced have been demonstrated with specific examples in \Cref{section : num exp}, but are applicable to arbitrary potential functions. A natural future step would thus be to apply these algorithms to problems of practical importance. Furthermore, while we have presented the general idea behind generalizing the circuit constructions to higher dimensions, the detailed complexity analysis and numerical experiments have not been investigated. Although the ancilla-assisted approach is suitable for many adaptive piecewise polynomial functions, we also anticipate the existence of an ancilla-free implementation for such functions based on adaptive grids, which can be investigated further. Finally, since the identification of a continuous variable quantum system from data requires, as a prerequisite, a parametrization of the infinite-dimensional space of Hamiltonians, it may be of interest to pursue continuous variable process tomography under the hypothesis of piecewise polynomial potentials.

\section*{Acknowledgements} Authors thank Sarah Mostame for several helpful discussions, and Dan Fortunato for sharing his  \MYhref{https://github.com/danfortunato/treefun}{\texttt{treefun}} repository which was used for creating adaptive discretizations used in this work. JS and SV acknowledge support from NSF under grant DMS-2038030. TA and JS acknowledge partial support from the Center for Computational Mathematics at the Flatiron Institute.
\bibliographystyle{unsrt}
\bibliography{references}

\onecolumn\newpage
\appendix

\section{Proof of Lemma \ref{lem:general}}\label{app. gen eig eq proof}
By the definition of $S(k)$ in \eqref{eq: labeling func}, the indicator function $\mathbbm{1}_{A_s}(x_k)$ satisfies
\begin{equation}
    \mathbbm{1}_{A_s}(x_k) =
    \begin{cases}
    1 & \text{for } S(k) = s\\
    0 & \text{otherwise.}
    \end{cases}
\end{equation}
This implies that
\begin{align}
    \mathbbm{1}_{A_s}(x_k) 
    & = \prod_{a=1}^{m}\frac{1+(-1)^{s_a \oplus S_a(k)}}{2} \\
    & = \frac{1}{M} \prod_{a=1}^{m} \big{[}1 + (-1)^{s_a} (-1)^{S_a(k)}\big{]} \\
    & = \frac{1}{M}\sum_{t=0}^{M-1} \prod_{a=1}^{m}(-1)^{s_a t_a} (-1)^{S_a(k) t_a} \\
    & = \frac{1}{M}\sum_{t=0}^{M-1} (-1)^{s\cdot t}\prod_{a=1}^{m } (-1)^{S_a(k) t_a}.
    \label{indicator function result}
\end{align}
By inserting \eqref{indicator function result} into \eqref{e:piecewise} and rearranging the summations, we arrive at \eqref{eq. general formula}.

\section{Hamiltonian representations in one dimension}\label{app. gen uniform ham rep proof}

First consider the ancilla-free implementation. \Cref{lem:general} and \eqref{e: uniform label} implies that for any computational basis state $\ket{k}$, 
\begin{equation}\label{eq. first step in der}
    f(x_k) \ket{k}= \sum^{M-1}_{t=0} g_{t}(x_k)\prod_{a=1}^{m}Z_a^{t_a}\ket{k},
\end{equation}
where $g_{t}(x)$ are quadratic polynomials for which the coefficients can be pre-computed. By applying the identity $k_a\ket{k} = \frac{1}{2}(1-Z_a)\ket{k}$, we can expand $x_k$ in terms of $k_a$ to arrive at
\begin{equation}\label{eq. quad pauli}
    g_{t}(x_k) \ket{k} = \bigg{(}\sum_{b=1}^{n}\sum_{c=b+1}^{n}A^{(t)}_{bc}Z_bZ_c + \sum_{b=1}^{n}B^{(t)}_{b}Z_b + C^{(t)} \bigg{)}\ket{k},
\end{equation}
where $A^{(t)}_{bc}$, $B^{(t)}_{b}$, and $C$ are constants obtained from the coefficients of $g_{t}(x)$. Inserting \eqref{eq. quad pauli} into \eqref{eq. first step in der}, we notice that the factor consisting of Pauli-Z operators in \eqref{eq. first step in der} is invariant under this multiplication, changing only the coefficients in $g_{t}(x_k)$, and we arrive at the desired form. Particularly, the change in the coefficients involves a permutation of the $t$ index.

Similarly, with the labels encoded in the ancillary quantum register, the equation that corresponds to \eqref{eq. first step in der} in the ancilla-assisted implementation is
\begin{equation}\label{eq. first step in der adap}
     f(x_k) \ket{k}= \sum^{M-1}_{t=0} g_{t}(x_k)\prod_{a=1}^{m}Z^{t_a}_{n+a} \ket{k}.
\end{equation}
Again, we insert \eqref{eq. quad pauli} into \eqref{eq. first step in der adap} to arrive at \eqref{eq. ham rep adaptive}. This shows that the difference between \eqref{e:h_f} and \eqref{eq. ham rep adaptive} originates from the index on the Pauli-Z operators.

\section{CNOT staircase form and related circuit constructions}\label{app. cnot staircase}

Some simple quantum circuits constructed based on the CNOT staircase form are shown in \Cref{fig. CNOT staircase 1}. We see that this construction corresponds to a single rotation gate in the single-qubit limit, and its generalization to more qubits can be achieved by repeatedly adding two more CNOT gates on both sides of the circuit. The CNOT staircase form is not limited to Pauli-Z strings, but can also be realized when different Pauli operators are included. In such cases, the Pauli-X or Pauli-Y operators can be handled by changing the basis using additional single-qubit gates, as illustrated by the additional Hadamard gates in \Cref{fig. CNOT staircase 2}.

\begin{figure}[t]
    \begin{subfigure}{\columnwidth}
        \centering
        $e^{-i\frac{\mu}{2}Z_{l_0}}$
         = 
        \begin{quantikz}
            \lstick{$q_{l_0}$} & \gate{R_{\text{z}}(\mu)} & \qw
        \end{quantikz}
        \label{fig. CNOT staircase 1a}
    \end{subfigure}
    
    \begin{subfigure}{\columnwidth}
    \centering
    $e^{-i\frac{\mu}{2}Z_{l_0}Z_{l_1}}$
     = 
    \begin{quantikz}
        \lstick{$q_{l_0}$} & \targ{} & \gate{R_{\text{z}}(\mu)} & \targ{} & \qw\\
        \lstick{$q_{l_1}$} & \ctrl{-1} & \qw & \ctrl{-1} & \qw
    \end{quantikz}
    \label{fig. CNOT staircase 1b}
    \end{subfigure}
    \begin{subfigure}{\columnwidth}
    \centering
    $e^{-i\frac{\mu}{2}Z_{l_0}Z_{l_1}Z_{l_2}}$
     = 
    \begin{quantikz}
        \lstick{$q_{l_0}$} & \qw & \targ{} & \gate{R_{\text{z}}(\mu)} & \targ{} & \qw & \qw\\
        \lstick{$q_{l_1}$} & \targ{} & \ctrl{-1} & \qw & \ctrl{-1} & \targ{} & \qw\\
        \lstick{$q_{l_2}$} & \ctrl{-1} & \qw & \qw & \qw  & \ctrl{-1} & \qw
    \end{quantikz}
    \label{fig. CNOT staircase 1c}
    \end{subfigure}
\mcaption{``CNOT staircase'' implementation of (top) single-, (middle) two-, and (bottom) three-qubit rotation-Z operations for distinct $l_0$, $l_1$, and $l_2$. The role of each qubit in these construction can be permuted freely, such that the rotation-Z gate is placed on a different qubit, and the target and control qubits of the CNOT gates are updated accordingly.}\label{fig. CNOT staircase 1}
\end{figure}

\begin{figure}[t]
  \centering
    $e^{-i\frac{\mu}{2}X_{l_0}Z_{l_1}Z_{l_2}}$
     = 
    \begin{quantikz}
        \lstick{$q_{l_0}$} & \gate{H} & \targ{} & \gate{R_{\text{z}}(\mu)} & \targ{} & \gate{H} & \qw\\
        \lstick{$q_{l_1}$} & \targ{} & \ctrl{-1} & \qw & \ctrl{-1} & \targ{} & \qw\\
        \lstick{$q_{l_2}$} & \ctrl{-1} & \qw & \qw & \qw  & \ctrl{-1} & \qw
    \end{quantikz}
    =
    \begin{quantikz}
        \lstick{$q_{l_0}$} & \qw & \control{} & \gate{R_{\text{x}}(\mu)} & \control{} & \qw & \qw\\
        \lstick{$q_{l_1}$} & \targ{} & \ctrl{-1} & \qw & \ctrl{-1} & \targ{} & \qw\\
        \lstick{$q_{l_2}$} & \ctrl{-1} & \qw & \qw & \qw  & \ctrl{-1} & \qw
    \end{quantikz}
    \mcaption{An example implementation of the ``CNOT staircase'' form when the Pauli-string that generates the unitary operation includes a Pauli-X operator, alongside Pauli-Z operators, for distinct $l_0$, $l_1$, and $l_2$.}
    \label{fig. CNOT staircase 2}
\end{figure}

\begin{figure}[t]
    \begin{subfigure}{\columnwidth}
        \centering
        $e^{-i\frac{\mu}{2}Z_{l_0}Z_{l_1}Z_{l_2}}$
     = 
    \begin{quantikz}
        \lstick{$q_{l_0}$} & \targ{} & \targ{} & \gate{R_{\text{z}}(\mu)} & \targ{} & \targ{} & \qw\\
        \lstick{$q_{l_1}$} & \qw & \ctrl{-1} & \qw & \ctrl{-1} & \qw & \qw\\
        \lstick{$q_{l_2}$} & \ctrl{-2} & \qw & \qw & \qw  & \ctrl{-2} & \qw
    \end{quantikz}
    \end{subfigure}
    \begin{subfigure}{\columnwidth}
    \centering
    $e^{-i\frac{\mu}{2}Z_{l_0}Z_{l_1}Z_{l_2}Z_{l_3}}$
     = 
    \begin{quantikz}
        \lstick{$q_{l_0}$} & \targ{} & \targ{} & \targ{} & \gate{R_{\text{z}}(\mu)} & \targ{} & \targ{} & \targ{} & \qw\\
        \lstick{$q_{l_1}$} & \qw & \qw & \ctrl{-1} & \qw & \ctrl{-1} & \qw & \qw & \qw\\
        \lstick{$q_{l_2}$} & \qw & \ctrl{-2} & \qw & \qw & \qw  & \ctrl{-2} & \qw & \qw\\
        \lstick{$q_{l_3}$} & \ctrl{-3} & \qw & \qw & \qw & \qw & \qw & \ctrl{-3} & \qw
    \end{quantikz}
    \end{subfigure}
    \mcaption{An example of a different construction that is equivalent to the CNOT staircase form in (top) three and (bottom) four qubits for distinct $l_0$, $l_1$, $l_2$, and $l_3$.}
    \label{fig. CNOT staircase 3}
\end{figure}

Note that the CNOT staircase form is not the only systematic construction of unitaries generated by tensor products of Pauli operators, and a different construction is illustrated in \Cref{fig. CNOT staircase 3}. In order to cancel as many neighboring CNOT gates as possible, we combine the two constructions in this work by applying the latter method to distinguish subdomains and the standard CNOT staircase form to implement the polynomials.

Whenever a linear combination of Pauli-Z strings that generates a desired unitary operation has a common factor of tensor products between two or more Pauli-Z operators, there are neighboring CNOT gates in the CNOT staircase that can be removed, as shown in \Cref{fig. CNOT staircase 4}. By grouping terms with the same indices together, we obtain such common factors, which enables the reduction of the circuit depth. For the other part of the quantum circuit, which distinguishes the subdomain, we apply the circuit construction shown in \Cref{fig. CNOT staircase 3}. With this circuit construction, the number of neighboring CNOT gates between $U^{(t_1)}$ and $U^{(t_2)}$ for $t_1 \neq t_2$  is maximized if we order $t$ in sequency order, which is given by the Gray code. An example of such cancellation is shown in \Cref{fig. CNOT staircase 5}.

\begin{figure}[t]
  \centering
    $e^{-i\frac{1}{2}(\mu_aZ_{l_0} + \mu_b)Z_{l_1}Z_{l_2}}$
     = $e^{-i\frac{\mu_a}{2}Z_{l_0}Z_{l_1}Z_{l_2}} e^{-i\frac{\mu_b}{2}Z_{l_1}Z_{l_2}}$
     =
     \begin{quantikz}
        \lstick{$q_{l_0}$} & \qw & \targ{} & \gate{R_{\text{z}}(\mu_a)} & \targ{} & \qw & \qw & \qw & \qw & \qw\\
        \lstick{$q_{l_1}$} & \targ{} & \ctrl{-1} & \qw & \ctrl{-1} & \targ{} & \targ{} & \gate{R_{\text{z}}(\mu_b)} & \targ{} &\qw\\
        \lstick{$q_{l_2}$} & \ctrl{-1} & \qw & \qw & \qw & \ctrl{-1} & \ctrl{-1} & \qw & \ctrl{-1} & \qw\\
    \end{quantikz}
    =
    \begin{quantikz}
        \lstick{$q_{l_0}$} & \qw & \targ{} & \gate{R_{\text{z}}(\mu_a)} & \targ{} & \qw & \qw\\
        \lstick{$q_{l_1}$} & \targ{} & \ctrl{-1} & \gate{R_{\text{z}}(\mu_b)} & \ctrl{-1} & \targ{} & \qw\\
        \lstick{$q_{l_2}$} & \ctrl{-1} & \qw & \qw & \qw & \ctrl{-1} & \qw\\
    \end{quantikz}
    \mcaption{An example of how to implement a unitary operator generated by a sum of Pauli-Z strings with a common factor, for distinct $l_0$, $l_1$, and $l_2$. The equality between the two quantum circuits can be identified by cancelling neighboring CNOT gates and using a commutation rule between the CNOT and rotation-Z gates.}
    \label{fig. CNOT staircase 4}
\end{figure}

\begin{figure}[t]
    \begin{subfigure}{\textwidth}
        \centering     $U^{(7)}U^{(6)}U^{(5)}U^{(4)}U^{(3)}U^{(2)}U^{(1)}U^{(0)}$
     = 
\par\medskip
\resizebox{0.95\textwidth}{!}{\begin{quantikz}
    \lstick{$q_{4\ldots n}$} & \gate{P_0} & \gate{P_1} & \gate{P_2} & \qw & \gate{P_3} & \qw & \gate{P_4} & \qw & \gate{P_5} & \qw & \qw & \gate{P_6} & \qw & \qw & \qw & \gate{P_7} & \qw & \qw & \qw\\
        \lstick{$q_{3}$} & \qw & \ctrl{-1} & \qw & \targ{} & \ctrl{-1} & \targ{} & \qw & \targ{} & \ctrl{-1} & \targ{} & \qw & \qw & \qw & \targ{} & \targ{} & \ctrl{-1} & \targ{} & \targ{} & \qw\\
        \lstick{$q_{2}$} & \qw & \qw & \ctrl{-2} & \ctrl{-1} & \qw & \ctrl{-1} & \qw & \qw & \qw & \qw & \targ{} & \ctrl{-2} & \targ{} & \qw & \ctrl{-1} & \qw & \ctrl{-1} & \qw & \qw\\
        \lstick{$q_{1}$} & \qw & \qw & \qw & \qw & \qw & \qw & \ctrl{-3} & \ctrl{-2} & \qw & \ctrl{-2} & \ctrl{-1} & \qw & \ctrl{-1} & \ctrl{-2} & \qw & \qw & \qw & \ctrl{-2} & \qw
    \end{quantikz}\unskip} 
    \end{subfigure}
\par\bigskip\vspace{0.3cm}
    \begin{subfigure}{\columnwidth}
    \centering
    $U^{(1)}U^{(5)}U^{(7)}U^{(3)}U^{(2)}U^{(6)}U^{(4)}U^{(0)}$
     = 
     \par\medskip
    \resizebox{0.88\textwidth}{!}{ \begin{quantikz}
    \lstick{$q_{4\ldots n}$} & \gate{P_0} & \gate{P_4} & \qw & \gate{P_6} & \qw & \gate{P_2} & \qw & \gate{P_3} & \qw & \gate{P_7} & \qw & \gate{P_5} & \qw & \gate{P_1} & \qw\\
        \lstick{$q_{3}$} & \qw & \qw & \qw & \qw & \qw & \qw & \targ{} & \ctrl{-1} & \targ{} & \ctrl{-1} & \targ{} & \ctrl{-1} & \targ{} & \ctrl{-1} & \qw\\
        \lstick{$q_{2}$} & \qw & \qw & \targ{} & \ctrl{-2} & \targ{} & \ctrl{-2} & \ctrl{-1} & \qw & \qw & \qw & \ctrl{-1} & \qw & \qw & \qw & \qw\\
        \lstick{$q_{1}$} & \qw & \ctrl{-3} & \ctrl{-1} & \qw & \ctrl{-1} & \qw & \qw & \qw & \ctrl{-2} & \qw & \qw & \qw & \ctrl{-2} & \qw & \qw
    \end{quantikz}\unskip}
    \end{subfigure}
    \mcaption{Implementing the same unitary operation in (top) Paley order and (bottom) sequency order. The latter can be used to reduce the circuit depth by cancelling neighboring CNOT gates. In the context of this work, the multi-qubit operations $P_t$ here are placeholders that correspond to the polynomial evaluations in the context of this work.}
    \label{fig. CNOT staircase 5}
\end{figure}

\section{Complexity of quantum circuits}\label{app. qc complexity}

In the following, we describe the gate counts of the ancilla-free and ancilla-assisted implementations when $\tau = 0$, in terms of $n$, $m$, and $l$. We restrict our attention to those piecewise functions that are non-trivial, i.e., $K \geq 2$.

\subsection{Ancilla-free Implementation}\label{app. qc complexity 1}

The ancilla-free implementation consists of rotation-Z gates and CNOT gates. The total number of rotation-Z gates, $\text{(RZ)}^{\text{a-f}}_{\text{tot}}$, can be decomposed into
\begin{equation}
    \text{(RZ)}^{\text{a-f}}_{\text{tot}} =  \sum^{M-1}_{t=0}\text{(RZ)}^{(t)}_{\theta} +  \text{(RZ)}^{(t)}_{\phi} +  \text{(RZ)}^{(t)}_{\lambda},
\end{equation}
where $\text{(RZ)}^{(t)}_{\theta}$,  $\text{(RZ)}^{(t)}_{\phi}$, and $\text{(RZ)}^{(t)}_{\lambda}$ denote the number of rotation-Z gates in $U^{(t)}_{b,c}(\theta)$, $U^{(t)}_{b}(\phi)$, and $U^{(t)}(\lambda)$, respectively. All the rotation-Z gates are used to apply rotations determined by the coefficients of the piecewise quadratic function, so their gate counts correspond to the number of rotation angles $\theta^{(t)}_{bc}$, $\phi^{(t)}_{b}$, and $\lambda^{(t)}$ in \eqref{e:h_f}. Thus, the number of rotation-Z gates with angles parameterized by $\theta^{(t)}_{bc}$, $\phi^{(t)}_{b}$, and $\lambda^{(t)}$ can be determined as
\begin{align}
    \text{(RZ)}^{(t)}_{\theta} &= \sum_{b=m+1}^{n}\sum_{c=b+1}^{n}1 = \frac{1}{2}(n-m)(n-m-1)\\
    \text{(RZ)}^{(t)}_{\phi} &= \sum_{b=m+1}^{n}1 = n-m\\
    \text{(RZ)}^{(t)}_{\lambda} &=
     \begin{cases}
    	0 & \text{if $t=0$}\\
        1 & \text{otherwise},
     \end{cases}
\end{align}
where $\text{(RZ)}^{(t)}_{\lambda} = 0$ for $t = 0$ because in that case the gate would produce an irrelevant global phase, and can be removed.

The CNOT gates can both be used to evaluate the polynomial and to distinguish the subdomains, and we denote the gate counts corresponding to these two purposes as $\text{(CX)}^{\text{a-f}}_{\text{P}}$ and $\text{(CX)}^{\text{a-f}}_{\text{D}}$, such that the total CNOT gate count satisfies
\begin{equation}
    \text{(CX)}^{\text{a-f}}_{\text{tot}} =  \text{(CX)}^{\text{a-f}}_{\text{P}} +  \text{(CX)}^{\text{a-f}}_{\text{D}}.
\end{equation}
Since $\text{(CX)}^{\text{a-f}}_{\text{P}}$ describes the CNOT gates associated with angles $\theta^{(t)}_{bc}$ and $\phi^{(t)}_{b}$ in the CNOT staircase form (described in \Cref{app. cnot staircase}), it can be expressed in terms of their rotation-Z gate counts as
\begin{equation}
    \text{(CX)}^{\text{a-f}}_{\text{P}} =
    	2\big{[}\text{(RZ)}^{\text{a-f}}_{\theta} + \text{(RZ)}^{\text{a-f}}_{\phi}\big{]} - 2(n-m),
    	\label{eq: af cx count P}
\end{equation}
where we defined $\text{(RZ)}^{\text{a-f}}_{\theta} := \sum_{t=0}^{M-1}\text{(RZ)}_{\theta}^{(t)}$ and $\text{(RZ)}^{\text{a-f}}_{\phi} := \sum_{t=0}^{M-1}\text{(RZ)}_{\phi}^{(t)}$, and the subtraction by $2(n-m)$ is again due to the fewer gates needed when $t=0$. As we optimize the circuit as described in \Cref{subsection : algorithm for 1D}, there is one CNOT gate per intersection between $U^{(t_1)}$ and $U^{(t_2)}$ for $t_1 \neq t_2$ and $t_1, t_2 >0$, such that (refer to \cite{welch2014} for the related discussion on Walsh-Fourier bases)
\begin{equation}
    \text{(CX)}^{\text{a-f}}_{\text{D}} = 2^m - 2.
\end{equation}

Therefore, after simplifying, the total number of rotation-Z gates is given by
\begin{equation}
    \text{(RZ)}^{\text{a-f}}_{\text{tot}} = 2^{m-1}(n-m)(n-m-1) + 2^{m}(n-m)+2^{m}-1 = O(M\log{(N/M)}^2) \label{eq: af rz count}
\end{equation}
and the total number of CNOT gates is given by
\begin{equation}
    \text{(CX)}^{\text{a-f}}_{\text{tot}} = 2^{m}(n-m)(n-m-1) + 2(2^{m}-1)(n-m)+2^{m}-2 = O(M\log{(N/M)}^2).
\end{equation}

\subsection{Ancilla-assisted Implementation}\label{app. qc complexity 2}

The main difference between the ancilla-assisted and ancilla-free implementations is that the former involves an additional labeling operation and its uncomputation. For labelling, we consider the example implementation discussed in \Cref{app. labeling gates} here to obtain explicit gate counts. The structure of the ancilla-assisted implementation is otherwise very similar to that of the ancilla-free implementation. In particular, the number of rotation-Z gates associated with polynomial evaluation,
\begin{equation}
    \text{(RZ)}^{\text{a-a}}_{\text{P}} =  \sum^{M-1}_{t=0}\text{(RZ)}^{(t)}_{\widetilde\theta} +  \text{(RZ)}^{(t)}_{\widetilde\phi} +  \text{(RZ)}^{(t)}_{\widetilde\lambda},
\end{equation}
can be again counted from the number of rotation angles $\widetilde{\theta}^{(t)}_{bc}$, $\widetilde{\phi}^{(t)}_{b}$, and $\widetilde{\lambda}^{(t)}$ in \eqref{eq. ham rep adaptive}, given by
\begin{align}
    \text{(RZ)}^{(t)}_{\widetilde\theta} &= \sum_{b=1}^{n}\sum_{c=b+1}^{n}1 = \frac{1}{2}n(n-1)\\
    \text{(RZ)}^{(t)}_{\widetilde\phi} &= \sum_{b=1}^{n}1 = n\\
    \text{(RZ)}^{(t)}_{\widetilde\lambda} &=
     \begin{cases}
    	0 & \text{if $t=0$}\\
        1 & \text{otherwise}.
     \end{cases}
\end{align}
Including the additional $\text{(RZ)}^{\text{a-a}}_{\text{L}} = 2^l - 1$ rotation-Z gates from the labeling procedure in \Cref{app. labeling gates}, we can obtain the total rotation-Z gate count from
\begin{equation}
    \text{(RZ)}^{\text{a-a}}_{\text{tot}} = \text{(RZ)}^{\text{a-a}}_{\text{P}} + \text{(RZ)}^{\text{a-a}}_{\text{L}}.
\end{equation}

The number of CNOT gates is also similar, with $\text{(CX)}^{\text{a-a}}_{\text{D}} = 2^m - 2$ and
\begin{equation}
    \text{(CX)}^{\text{a-a}}_{\text{P}} =
    	2\big{[}\text{(RZ)}^{\text{a-a}}_{\widetilde\theta} + \text{(RZ)}^{\text{a-a}}_{\widetilde\phi}\big{]} - 2n, \label{eq: aa cx count P}
\end{equation}
for $\text{(RZ)}^{\text{a-a}}_{\theta} := \sum_{t=0}^{M-1}\text{(RZ)}_{\widetilde\theta}^{(t)}$ and $\text{(RZ)}^{\text{a-a}}_{\phi} := \sum_{t=0}^{M-1}\text{(RZ)}_{\widetilde\phi}^{(t)}$. Again, the total gate count can be obtained by including $\text{(CX)}^{\text{a-a}}_{\text{L}} = 2^{l+1}-4$ additional CNOT gates in
\begin{equation}
    \text{(CX)}^{\text{a-a}}_{\text{tot}} =  \text{(CX)}^{\text{a-a}}_{\text{P}} +  \text{(CX)}^{\text{a-a}}_{\text{D}} +  \text{(CX)}^{\text{a-a}}_{\text{L}}.
\end{equation}

The labeling procedure also consists of controlled-Z and rotation-X gates. Thus, the total gate count for each type of quantum gate is
\begin{align}
    \text{(RZ)}^{\text{a-a}}_{\text{tot}} &= 2^{m-1}n (n -1) + 2^{m}n +2^{m}+2^{l}-2 = O(M\log{(N)}^2)\label{eq: aa rz count}\\
    \text{(RX)}^{\text{a-a}}_{\text{tot}} &= \text{(RX)}^{\text{a-a}}_{\text{L}} = 2^{l+1} m = O(L\log{(M)})\\
    \text{(CZ)}^{\text{a-a}}_{\text{tot}} &= \text{(CZ)}^{\text{a-a}}_{\text{L}} = 4(2^l-1)m = O(L\log{(M)})\\
    \text{(CX)}^{\text{a-a}}_{\text{tot}} &= 2^{m}n (n-1) + 2(2^{m}-1)n +2^{m}+2^{l+1}-6 = O(M\log{(N)}^2)
\end{align}

\subsection{Applying the ancilla-assisted implementation on uniform piecewise quadratic functions} \label{app. qc complexity 3}

While in general the number of subdomains is less than or equal to the number of cells ($K \leq L$), uniform piecewise functions, as introduced in \Cref{section : classical preliminaries}, satisfy $K = L = M$. Even though the ancilla-free implementation is intuitively the most natural implementation of such functions, it is also possible to consider applying the ancilla-assisted approach. We show here, however, that the ancilla-free approach is always more efficient than the ancilla-assisted approach both in terms of space and time for uniform piecewise quadratic functions, and we expect this to be true also for other values of $\alpha$.

The total gate count $G^{\text{a-a}}_{\text{tot}}$ for the ancilla-assisted implementation is generally given by
\begin{equation}
    G^{\text{a-a}}_{\text{tot}} = \text{(RZ)}^{\text{a-a}}_{\text{tot}} + \text{(CX)}^{\text{a-a}}_{\text{tot}} + \text{(RX)}^{\text{a-a}}_{\text{tot}} +
    \text{(CZ)}^{\text{a-a}}_{\text{tot}},
    \label{eq: aa total count}
\end{equation}
and that for the ancilla-free implementation is
\begin{equation}
    G^{\text{a-f}}_{\text{tot}} = \text{(RZ)}^{\text{a-f}}_{\text{tot}} + \text{(CX)}^{\text{a-f}}_{\text{tot}}.
    \label{eq: af total count}
\end{equation}
Focusing on the rotation-Z gate counts in \eqref{eq: af rz count} and \eqref{eq: aa rz count}, we can easily see that
\begin{align}
    &\text{(RZ)}^{\text{a-a}}_{\text{tot}} - \text{(RZ)}^{\text{a-f}}_{\text{tot}}\nonumber\\
    =& \big{[}2^{m-1}n (n -1) + 2^{m}n +2^{m}+2^{l}-2\big{]} -  \big{[}2^{m-1}(n-m)(n-m-1) + 2^{m}(n-m)+2^{m}-1\big{]}\\
    =& 2^{m-1}m(2n-m) + 2^l - 1\\
    >& 0,
\end{align}
where in the last step we used the fact that $n \geq l \geq m \geq 1$ for piecewise functions with at least one knot. We can similarly compare the CNOT gates, using \eqref{eq: af cx count P} and \eqref{eq: aa cx count P}, to obtain \begin{align}
    &\text{(CX)}^{\text{a-a}}_{\text{tot}} - \text{(CX)}^{\text{a-f}}_{\text{tot}}\nonumber\\
    =& \big{[}\text{(CX)}^{\text{a-a}}_{\text{P}} +  \text{(CX)}^{\text{a-a}}_{\text{D}} +  \text{(CX)}^{\text{a-a}}_{\text{L}}\big{]} - \big{[}\text{(CX)}^{\text{a-f}}_{\text{P}} +  \text{(CX)}^{\text{a-f}}_{\text{D}}\big{]}\\
    =& \text{(CX)}^{\text{a-a}}_{\text{P}} +  \text{(CX)}^{\text{a-a}}_{\text{L}} - \text{(CX)}^{\text{a-f}}_{\text{P}}\\
    \geq& \text{(CX)}^{\text{a-a}}_{\text{P}} - \text{(CX)}^{\text{a-f}}_{\text{P}}\\
    =& \big{[}2^mn(n-1)+2^{m+1}n-2n\big{]} - \big{[}2^m(n-m)(n-m-1)+2^{m+1}(n-m)-2(n-m)\big{]}\\
    =& 2m\big{[}2^{m-1}(2n-m) - 1\big{]}\\
    >& 0.
\end{align}
Thus, we obtain
\begin{equation}
    G^{\text{a-a}}_{\text{tot}} - G^{\text{a-f}}_{\text{tot}}
    \geq \big{[}\text{(RZ)}^{\text{a-a}}_{\text{tot}} + \text{(CX)}^{\text{a-a}}_{\text{tot}}\big{]} - \big{[}\text{(RZ)}^{\text{a-f}}_{\text{tot}} + \text{(CX)}^{\text{a-f}}_{\text{tot}}\big{]}
    > 0,
\end{equation}
which shows that the total gate count with the ancilla-assisted implementation is always greater than that of the ancilla-free implementation for a uniform piecewise quadratic function. Note that we arrived at this result without even considering the additional gates that arise from labeling. Since the ancilla-assisted implementation also introduces an additional $m$-qubit quantum register, we conclude that in this case it is always advantageous to apply the ancilla-free implementation from the viewpoints of both space and time. This confirms our intuition, since in this case we are just mapping the $m=l$ most significant bits of the main register, which together defines the cells, to the ancillary label register with the same number of qubits.

\subsection{Applying the ancilla-free implementation on non-uniform piecewise quadratic functions} \label{app. qc complexity 4}

For non-uniform piecewise functions, we can similarly apply both the ancilla-free and ancilla-assisted implementations. When applying the ancilla-assisted implementation, we encode the labels of the $K$ non-uniform subdomains into an ancillary label register. When applying the ancilla-free implementation, however, we view the non-uniform piecewise function defined by subdomains \eqref{def. ada subdomain} instead as uniform piecewise functions on subdomains corresponding to the underlying cells in \eqref{def. uni subdomain}. The gate count would then correspond to that in \Cref{app. qc complexity 1}, where the number of subdomains would be understood as $L$. We show here that the ancilla-free implementation is no longer always more advantageous compared to the ancilla-assisted implementation in this case, and the relative efficiencies of the two methods depend on the values of $n$, $m$, and $l$.

As $m = l$ in the ancilla-free implementation, we obtain the total number of rotation-Z gates
\begin{equation}
    \text{(RZ)}^{\text{a-f}}_{\text{tot}} = 2^{l-1}(n-l)(n-l-1) + 2^{l}(n-l)+2^{l}-1
\end{equation}
and the total number of CNOT gates
\begin{equation}
    \text{(CX)}^{\text{a-f}}_{\text{tot}} = 2^{l}(n-l)(n-l-1) + 2(2^{l}-1)(n-l)+2^{l}-2.
\end{equation}
Since the total gate count of each of the two implementations is described by \eqref{eq: aa total count} and \eqref{eq: af total count}, we obtain that
\begin{align}
    &G^{\text{a-a}}_{\text{tot}} - G^{\text{a-f}}_{\text{tot}}\nonumber\\
    =& \big{[}2^{m-1}n (n -1) + 2^{m}n +2^{m} + 2^{l} - 2 + 2^{m}n (n-1) + 2(2^{m}-1)n +2^{m}+2^{l+1}-6 + 2^{l+1} m + 4(2^l-1)m\big{]}\nonumber\\
    &- \big{[}2^{l-1}(n-l)(n-l-1) + 2^{l}(n-l)+2^{l}-1 + 2^{l}(n-l)(n-l-1) + 2(2^{l}-1)(n-l)+2^{l}-2\big{]}\\
    =& A_G n^2 + B_G n + C_G,
\end{align}
where
\begin{align}
    A_G &= 3 (2^{m - 1} - 2^{l - 1})\\
    B_G &= 3(2^{m-1} - 2^{l-1} + 2^l l)\\
    C_G &= -3\times 2^{l-1}l^2+3\times 2^{l+1}m + 3 \times 2^{l-1}l + 2^l- 2l + 2^{m+1}-4m-5.
\end{align}
Thus, we observe that the difference between the two gate counts is a quadratic function in $n$. By solving the quadratic inequality corresponding to $G^{\text{a-a}}_{\text{tot}} < G^{\text{a-f}}_{\text{tot}}$, we obtain values of $n$ (expressed in terms of $m$ and $l$) such that the ancilla-assisted implementation can reduce the total gate count compared to the ancilla-free implementation. We illustrate this scenario with the Eckart barrier example in \Cref{fig: gate count}, where the intersections between the two lines indicate the critical point (the other solution to the quadratic equation always lies below $l$ in our numerical experiments), above which the ancilla-assisted implementation would be advantageous in terms of gate count.

\section{Labeling operation with single- and two-qubit gates}\label{app. labeling gates}
Labeling generally non-uniform subdomains can be done using at most $O(L\log{(M)})$ multi-controlled X gates, where the control qubits are at most $l$ qubits in the main register. In the cases where $l  = 1$ or $l  = 2$, a multi-controlled X gate simply corresponds to a single CNOT or Toffoli gate, and it can also be decomposed into Toffoli, CNOT, and single-qubit gates either with or without an additional quantum register when $l \geq 3$. If no ancilla qubits are used, each multi-controlled X operation can be performed with $O(\log{(L)}^2)$ Toffoli, CNOT, and single-qubit gates \cite{nielsen2011}. Otherwise, we can introduce an ancilla qubit to perform the operation with $O(\log{(L)})$ Toffoli gates. Such an implementation therefore introduces $O(L\log{(L)}^2\log{(M)})$ operations without ancilla qubits and $O(L\log{(L)}\log{(M)})$ operations with a single ancilla qubit, using Toffoli, CNOT, and single-qubit gates. To improve upon this, we introduce here an implementation that uses only $O(L\log{(M)})$ single- and two- qubit gates without utilizing ancilla qubits. Note that the method presented here is not necessarily optimal, and it is possible that ideas from existing works (see, e.g., \cite{barenco1995elementary, liu2008analytic, lanyon2009simplifying, saeedi2013linear, iten2016quantum, he2017decomposition, luo2016comment, bae2020quantum, dasilva2022linear}) can provide improvements. Nevertheless, we introduce our implementation here as an explicit example.

First consider decomposing a single multi-controlled X gate, where there are $l$ controls either conditioned on 0 or 1 (we sandwich the control with a pair of X gates when conditioning on 0), and a single X operation that acts on target qubit $b \in \{1,2,\ldots, m\}$ on the label register. Since in the end we want to map a subset of indices $\Gamma^{(S(k))}$ into a single index $S(k)$, we denote the controls with bit strings defined by the binary expansion of some $\tilde{r} \in \Gamma^{(S(k))}$, where $\tilde{r} \in \{0,1,2,\ldots,L-1\}$ and $\tilde{r}_a = k_a$ for $a=1,2,\ldots,l$. We can then use, without ambiguity, the notation $S(\tilde{r})$ instead of $S(k)$ to emphasize the $\tilde{r}$-dependence of the labels. In terms of Pauli operators, we can write the unitary operator corresponding to the multi-controlled X gate with controls defined by $\tilde{r}$ and target qubit $b$ as
\begin{align}
    U^{(\tilde{r}, b)}_{\text{MCX}} &= e^{i\frac{\pi}{2}\prod^{l }_{a=1}\big{[}\frac{1}{2}(I_a+(-1)^{\tilde{r}_a}Z_{a})\big{]}(I_b-X_{b})}\\
    &= \prod^{L-1}_{\tilde{t}=0}e^{i\frac{\pi}{2L}(-1)^{\tilde{r}\cdot \tilde{t}}\prod^{l}_{a=1}Z^{\tilde{t}_a}_{a}(I_b-X_b)}\\
    &=\prod^{L-1}_{\tilde{t}=0}e^{i\frac{\pi}{2L}(-1)^{\tilde{r}\cdot \tilde{t}}\prod^{l}_{a=1}Z^{\tilde{t}_a}_{a}}e^{-i\frac{\pi}{2L}(-1)^{\tilde{r}\cdot \tilde{t}}\prod^{l}_{a=1}Z^{\tilde{t}_a}_{a}X_b}.
    \label{eq. multi-controlled X decomposition}
\end{align}
Again, by considering the CNOT staircase form (\Cref{app. cnot staircase}) and viewing the product in $\tilde{t}$ as being sorted in sequency order, this expression corresponds to a quantum circuit that implements the multi-controlled X gate in $O(L)$ rotation-Z, rotation-X, controlled-Z and CNOT gates (see  \Cref{fig. CNOT staircase 2} and \Cref{fig. CNOT staircase 5}). In particular, the rotation-Z and rotation-X gates would each have angles given by $-\beta_{\tilde{r},\tilde{t}}$ and $+\beta_{\tilde{r},\tilde{t}}$, respectively, where $\beta_{\tilde{r},\tilde{t}} := \frac{\pi}{L}(-1)^{\tilde{r}\cdot \tilde{t}}$.

To map a set of bit strings in the main register to a single bit string in the label register, we need to convert the above multi-controlled X gate into a multi-controlled multi-X gate. The circuit that we have obtained through \eqref{eq. multi-controlled X decomposition} can be easily adapted to a multi-target gate by applying multiple multi-controlled X gates, one after another, with the exact same controls but applied onto different qubits in the label register. A multi-controlled multi-X gate with the target qubits corresponding to the bit string of $S(\tilde{r}) \equiv S(k)$ can thus be written as
\begin{align}
    U^{(\tilde{r})}_{\text{MCMX}} &= \prod_{b = 1}^{m}\big{(}U^{(\tilde{r}, b)}_{\text{MCX}}\big{)}^{S_b(\tilde{r})}\\
    &=\prod^{L-1}_{\tilde{t}=0}e^{i\frac{1}{2}\beta_{\tilde{r},\tilde{t}}\prod^{l}_{a=1}Z^{\tilde{t}_a}_{a}\sum_{b=1}^{m}S_b(\tilde{r})(I_{b}-X_{b})}\\
    &= \prod^{L-1}_{\tilde{t}=0}e^{i\frac{1}{2}\beta_{\tilde{r},\tilde{t}}\eta(\tilde{r})\prod^{l}_{a=1}Z^{\tilde{t}_a}_{a}}e^{-i\frac{1}{2}\beta_{\tilde{r},\tilde{t}}\prod^{l}_{a=1}Z^{\tilde{t}_a}_{a}\sum_{b=1}^{m}S_b(\tilde{r})X_b}
    \label{eq. multi-controlled multi-X decomposition}
\end{align}
where in the final step we defined the Hamming weight $\eta(\tilde{r}) := \sum_{b=1}^{m}S_b(\tilde{r})$ of the label corresponding to $\tilde{r}$, with the subscript $b$ denoting the $b$th bit in the binary expansion of $S(\tilde{r})$. The first exponential factor ensures that the relative phases between computational basis states are preserved, and the second factor builds up bit flips from X rotations. The summation over $b$ in \eqref{eq. multi-controlled multi-X decomposition} corresponds to applying the rotation-X gates in parallel onto each qubit $b$ that satisfies $S_b(\tilde{r}) = 1$, so that even though the number of gates increases compared to \eqref{eq. multi-controlled X decomposition}, the circuit depth remains unchanged.

Other than flipping multiple qubits in the label register, we also need to perform the multi-controlled multi-X gate with different bit strings on the control part to associate all $L$ uniform subdomains into their corresponding label. Similarly as the multi-target case, we can simply implement the multi-controlled multi-X gates one after another to obtain
\begin{align}\label{eq. labeling unitary decomposition}
    U_{S} &= \prod_{\tilde{r} = 0}^{L-1} U^{(\tilde{r})}_{\text{MCMX}}\\
    &= \prod_{\tilde{r} = 0}^{L-1}\prod^{L-1}_{\tilde{t}=0}e^{i\frac{1}{2}\beta_{\tilde{r},\tilde{t}}\eta(\tilde{r})\prod^{l}_{a=1}Z^{\tilde{t}_a}_{a}}e^{-i\frac{1}{2}\beta_{\tilde{r},\tilde{t}}\prod^{l}_{a=1}Z^{\tilde{t}_a}_{a}\sum_{b=1}^{m}S_b(\tilde{r})X_b}\\
    &= \prod^{L-1}_{\tilde{t}=0}e^{-i\frac{1}{2}\chi_{\tilde{t}}\prod^{l}_{a=1}Z^{\tilde{t}_a}_{a}}e^{-i\frac{1}{2}\sum_{b = 1}^{m}\xi_{\tilde{t},b}\prod^{l}_{a=1}Z^{\tilde{t}_a}_{a}X_b},
\end{align}
where the rotation angles are given by $\chi_{\tilde{t}} := -\sum_{\tilde{r} = 0}^{L-1}\beta_{\tilde{r},\tilde{t}}\eta(\tilde{r})$ and $\xi_{\tilde{t},b} := \sum_{\tilde{r} = 0}^{L-1}\beta_{\tilde{r},\tilde{t}}S_b(\tilde{r})$. One can observe that the structure of the quantum circuit is unchanged compared to the individual multi-controlled multi-X gates, since we were able to cancel neighboring CNOT gates and merge single-qubit rotation gates as we turned the product over $\tilde{r}$ into summations on the exponents. This indicates that $U_S$ can be implemented with the same circuit depth as a single multi-controlled X gate decomposed by \eqref{eq. multi-controlled X decomposition}.

\begin{figure}[t]
\centering
\begin{subfigure}{0.47\columnwidth}
    \centering
    \includegraphics[width=\columnwidth]{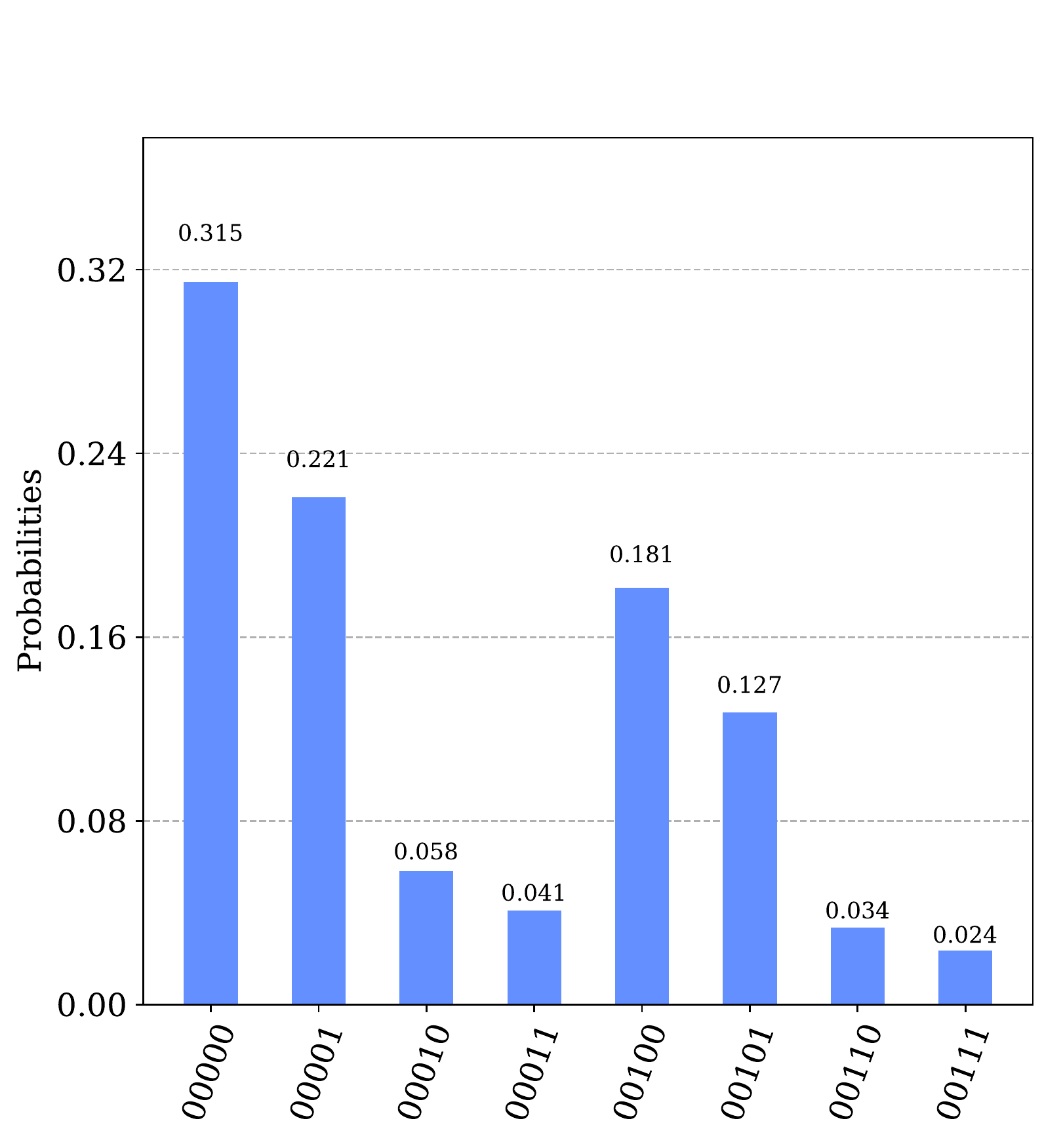}
\end{subfigure}\hspace{0.8cm}%
~
\begin{subfigure}{0.47\columnwidth}
    \centering
    \includegraphics[width=\columnwidth]{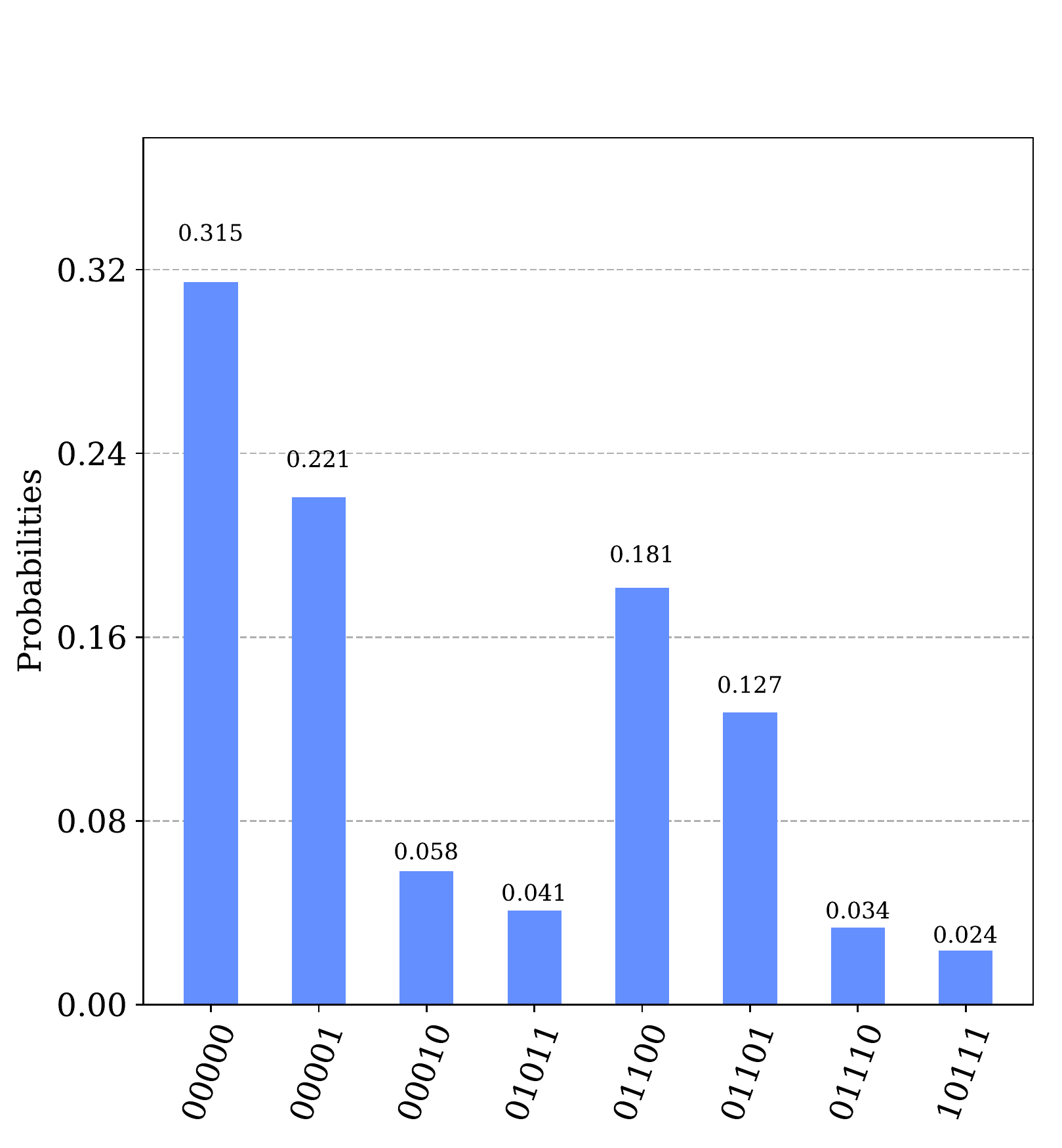}
\end{subfigure}
\begin{subfigure}{\columnwidth}
    \centering
    \includegraphics[width=0.98\columnwidth]{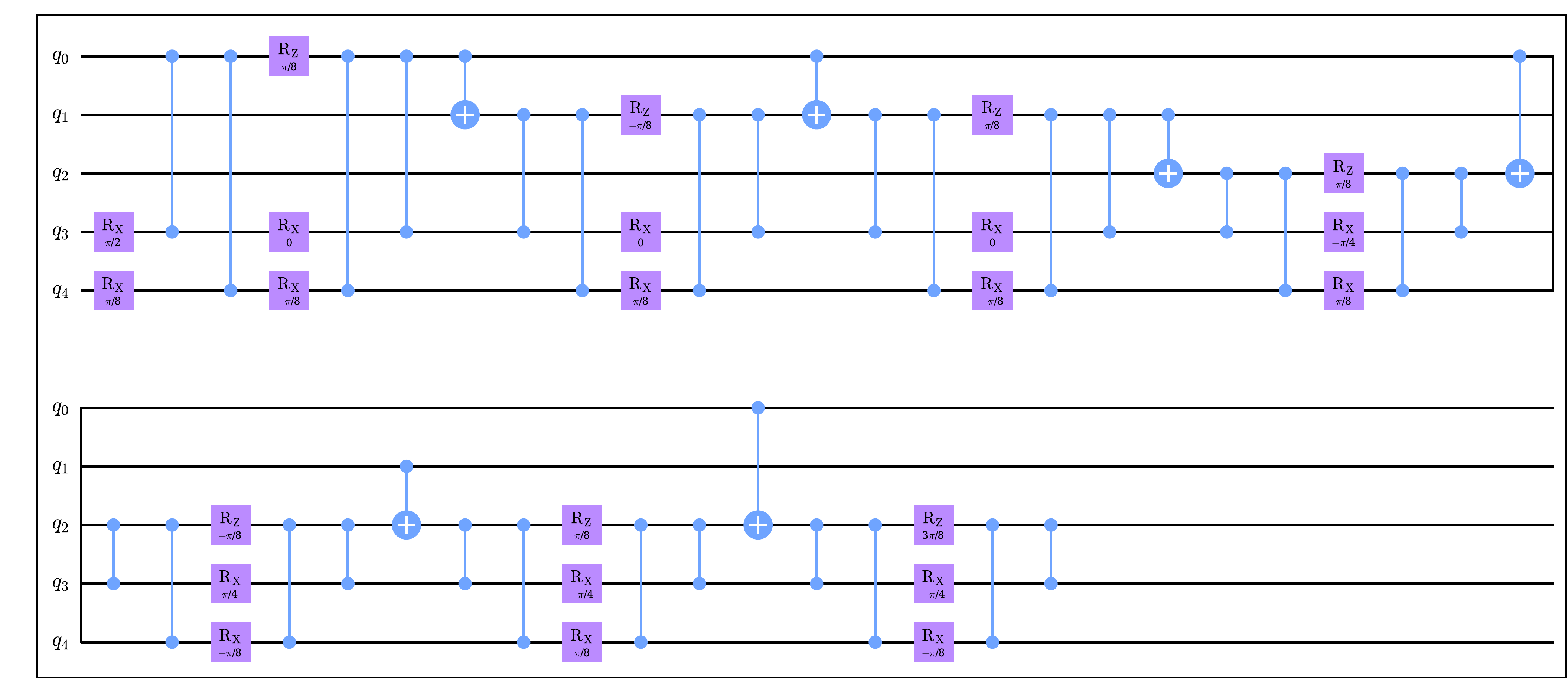}
\end{subfigure}
  \mcaption{Example implementation of the labeling operation introduced in \Cref{app. labeling gates}, where we encode $S(\tilde{r})$ onto $m=2$ qubits based on the value of $\tilde{r}$ in $l=3$ qubits. The example considers $\Gamma^{(00)} = \{000,001,010\}$, $\Gamma^{(01)} = \{011,100,101,110\}$, and $\Gamma^{(10)} = \{111\}$. (Top) Histograms obtained from measuring the arbitrarily-initialized quantum state in the computational basis (left) before and (right) after applying labeling gates. (Bottom) The quantum gates that map the labels from qubits $q_0, q_1$, and $q_2$ into qubits $q_3$ and $q_4$. The gates used for the initialization to the arbitrary superposition state is not shown.}
  \label{fig: labeling example}
\end{figure}

As an example, we perform the labeling operation controlled by $l = 3$ qubits and targeted to $m = 2$ qubits for $\Gamma^{(00)} = \{000,001,010\}$, $\Gamma^{(01)} = \{011,100,101,110\}$, and $\Gamma^{(10)} = \{111\}$, where we expressed $\tilde{r} \in \{0,1,2,\ldots,L-1\}$ and $S(\tilde{r}) \in \{0,1,2,\ldots,M-1\}$ in binary representation. We first initialize the $l = 3$ qubits to an arbitrary superposition state, for which the histogram obtained by measuring in the computational basis is shown on the top left of \Cref{fig: labeling example}. The quantum circuit that encodes the labels into the $m = 2$ qubits is shown on the bottom of the same figure, and the resulting histogram of the computational-basis measurement is shown on the top right.

With this method, we can perform the labeling operation with $2^l - 1$ rotation-Z, $2^l m$ rotation-X, $2(2^{l}-1)m$ controlled-Z, and $2^{l }-2$ CNOT gates, giving a total gate count of $3\times 2^{l }m+2^{l+1}-2m-3 = O(L\log{(M)})$. The quantum circuit is its own inverse operation, such that the total gate count doubles if we include the uncomputation step, which is required when applying the ancilla-assisted method that we introduce in this article to DQS.

\section{Different expression of Lemma \ref{lem:general} for the ancilla-free method in $d$-dimensions}\label{app. lemma for uniform}
For multivariate piecewise functions defined in $d$ dimensions, we can consider each dimension separately and introduce $K^{(i)}$ as the number of subdomains in dimension $i \in \{1,2,\ldots,d\}$, such that $K = \prod_{i=1}^{d}K^{(i)}$. We can then write the multivariate polynomial function as
\begin{align}
f(x) &= \sum_{s=0}^{K-1} f_{s}(x)\mathbbm{1}_{A_{s}}(x)\\
     &= \sum^{K^{(i)}-1}_{\substack{s^{(i)}=0\\ \forall \,1\leq i \leq d}} f_{s}(x)\mathbbm{1}_{A_{s}}(x)\\
    &\equiv \sum^{K^{(1)}-1}_{s^{(1)}=0}\sum^{K^{(2)}-1}_{s^{(2)}=0}\cdots \sum^{K^{(d)}-1}_{s^{(d)}=0} f_{s}(x)\mathbbm{1}_{A_{s}}(x),
\end{align}
where $s$ is to be understood in $d$ dimensions as $s \in [0: K^{(1)})\times \cdots \times [0: K^{(d)})$. With this, we obtain the following corollary to \Cref{lem:general}:
\begin{corollary}\label{cor:uniform D-dim}
For all $i \in \{1,2,\ldots,d\}$, for each $k^{(i)} \in \{0,1\}^{n^{(i)}}$, let
\begin{equation}
    S(k^{(i)}) := \sum_{s^{(i)}=0}^{K^{(i)}-1}s^{(i)}\mathbbm{1}_{A^{(i)}_s}(x_k^{(i)})
\end{equation}
denote the index of the cell containing the mesh point $x_k^{(i)} \in A_{S(k^{(i)})}^{(i)}$.
Then for all $k^{(i)} \in \{0,1\}^{n^{(i)}}$,
\begin{subequations}\label{eq. general uniform D-dim formula}
 \begin{equation}
     f(x_k) = \sum^{M^{(i)}-1}_{\substack{t^{(i)}=0\\ \forall \,1\leq i \leq d}} g_{t}(x_k)\prod_{i=1}^{d}\prod_{a=1}^{m ^{(i)}}(-1)^{S_a(k^{(i)})t^{(i)}_a},
     \label{eq. general uniform D-dim formula 1}
 \end{equation}
 where
 \begin{equation}
     g_{t}(x) \equiv \frac{1}{\prod_{i=1}^{d}M^{(i)}} \sum^{K^{(i)}-1}_{\substack{s^{(i)}=0\\ \forall \,1\leq i \leq d}}(-1)^{\sum_{i=1}^{d}s^{(i)}\cdot t^{(i)}}f_{s}(x).
     \label{eq. general uniform D-dim formula 2}
 \end{equation}
 \end{subequations}
\end{corollary}
A derivation of the quantum circuit for this $d$-dimensional case can be performed beginning with \Cref{cor:uniform D-dim}, and proceeding analogously to the one-dimensional case in \Cref{subsection : algorithm for 1D}.

\section{Piecewise quadratic approximation of cosine potential with $\epsilon = 10^{-1}$}\label{app. e.g. of num exp with m =2}

\begin{figure}[t]
\centering
\begin{subfigure}{0.47\columnwidth}
    \centering
    \includegraphics[width=\columnwidth]{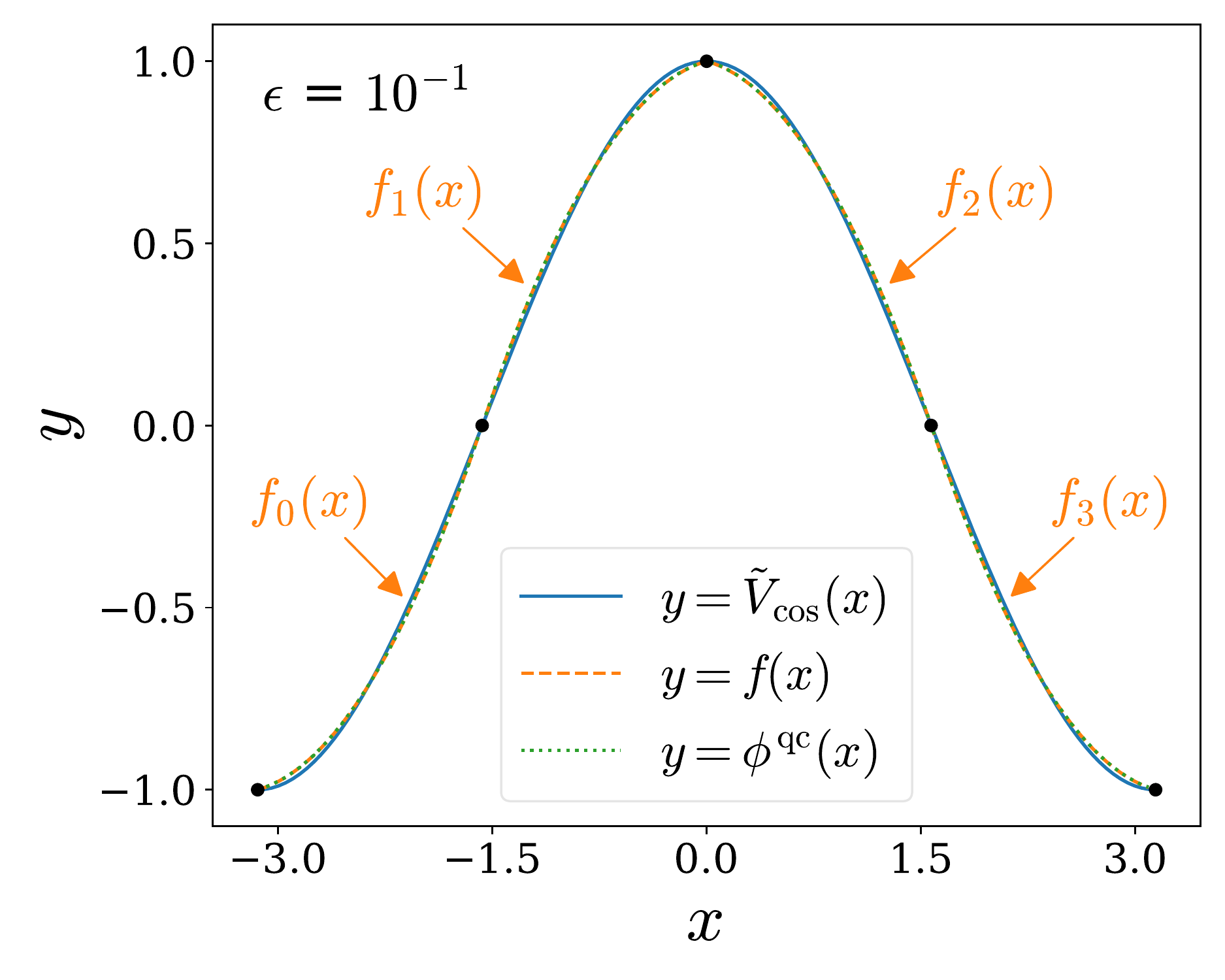}
\end{subfigure}\hspace{0.8cm}%
~
\begin{subfigure}{0.47\columnwidth}
    \centering
    \includegraphics[width=\columnwidth]{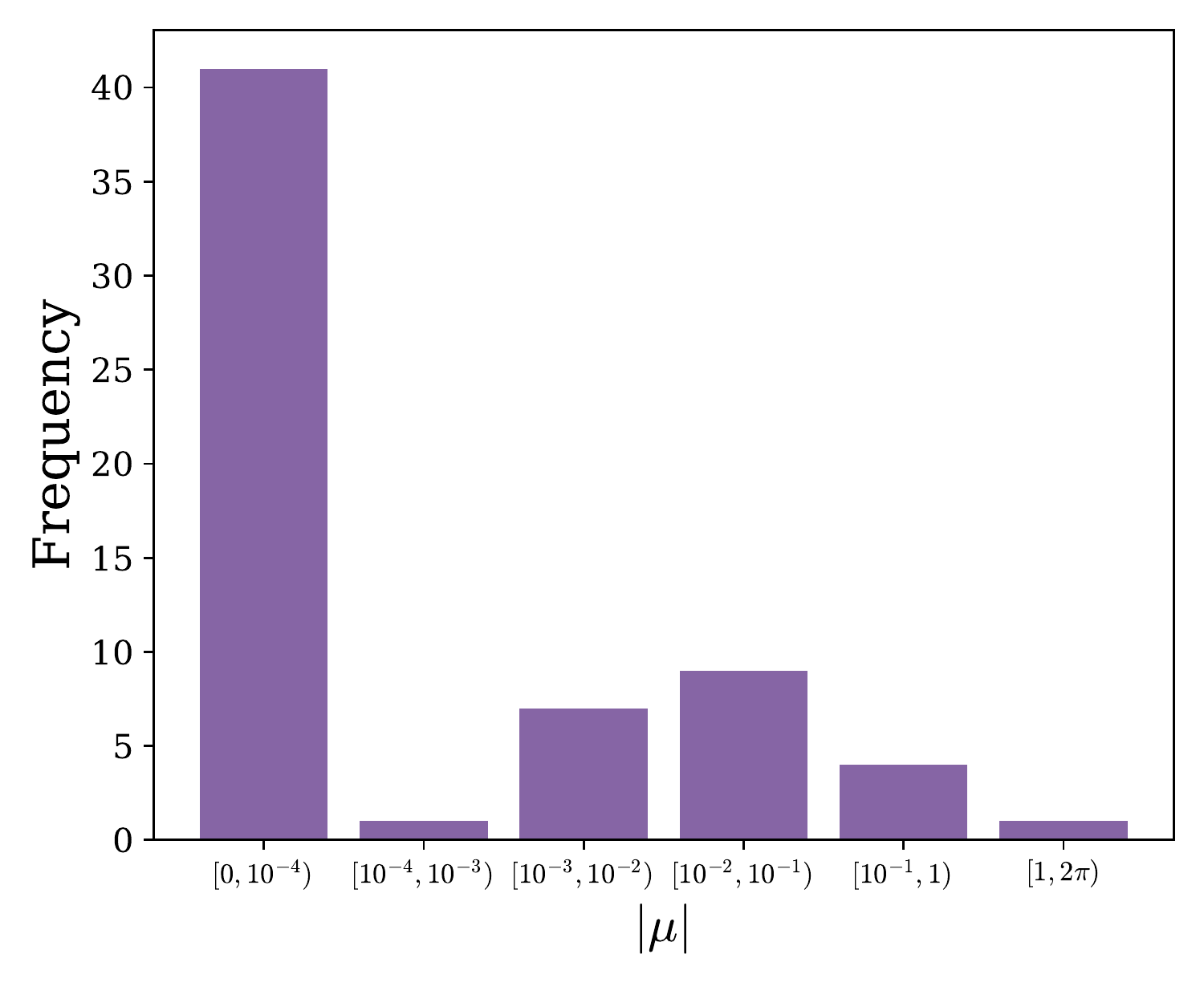}
\end{subfigure}
\begin{subfigure}{\columnwidth}
    \centering
    \includegraphics[width=0.985\columnwidth]{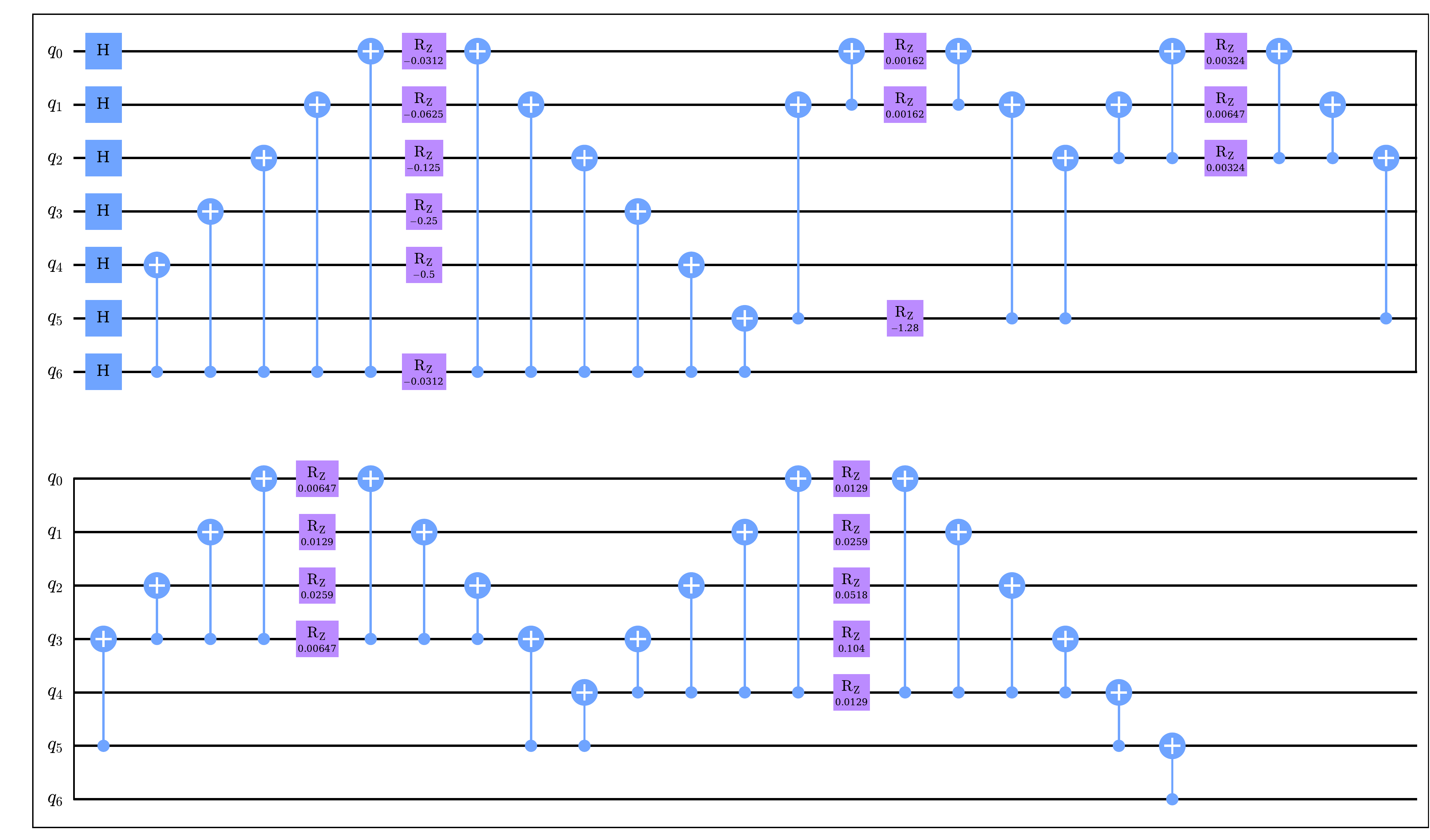}
\end{subfigure}
  \mcaption{Analysis of the implementation of the cosine function with piecewise approximations, where we choose $\epsilon = 10^{-1}$. (top left) The cosine function $\tilde{V}_{\text{cos}}$, its piecewise quadratic approximation function $f^{\epsilon}$, and the actual phase implemented by the quantum algorithm $\phi^{\text{qc}}$ with $\tau = 10^{-3}$ evaluated on $x_k$. Black dots indicate the positions of knots that connect the four quadratic functions. (top right) Histogram counting the number of times the absolute values of angles of rotation-Z gates, denoted as $\mu$, fall into the intervals $[0, 10^{-4})$, $[10^{-4}, 10^{-3})$, $[10^{-3}, 10^{-2})$, $[10^{-2}, 10^{-1})$, $[10^{-1}, 1)$, and $[1, 2\pi)$. The result justifies that the removal of rotation-Z gates with small angles and its surrounding CNOT gates can result in a significant reduction in the total gate count. (bottom) Quantum circuit constructed in the numerical experiment in \Cref{subsubsection : num exp results cos}, where $\epsilon = 10^{-1}$ and $\tau = 10^{-3}$. While the circuit depth is nearly halved compared to the $\tau=0$ circuit (not shown), the increase in $\delta(\tau)$ is negligible.}
  \label{fig: explicit example for ns=2}
\end{figure}

For a concrete example of the piecewise quadratic function in \Cref{subsubsection : num exp results cos}, consider the case where $\epsilon = 10^{-1}$. In this case, the classical algorithm generates a piecewise uniform function with $4$ subdomains, where the cosine function is being approximated by quadratic polynomials in each of them, as shown in the top left side of \Cref{fig: explicit example for ns=2}. The figure also shows a histogram of the absolute values of all angles involved in the quantum circuit when $\tau = 0$ (top right), from which we can observe that most angles are small in this case, justifying the effectiveness of removing rotations with angles smaller than some finite $\tau$. When implemented exactly by setting $\tau = 0$, the gate counts for rotation-Z gates and CNOT gates is 63 and 112, respectively. If we instead choose $\tau = 10^{-3}$, the gate counts are reduced, with only 21 rotation-Z gates and 40 CNOT gates. Moreover, the value of $\delta(\tau)$ is approximately 0.02343 for $\tau=0$ and 0.02382 for $\tau=10^{-3}$, which shows that such removal of gates only result in a small increase in $\delta(\tau)$, as can also be observed from \Cref{fig: data for cos ex}. The quantum circuit corresponding to this case is given in the bottom of \Cref{fig: explicit example for ns=2}. By using a classical simulator of a quantum computer, we can extract the actual phase $\phi^{\text{qc}}$ (adjusted by removing global phase) that was applied instead of $f^{\epsilon}$ due to the finite choice of $\tau$, and it is plotted alongside $\tilde{V}_{\text{cos}}$ and $f^{\epsilon}$ in the top left side of \Cref{fig: explicit example for ns=2}.

\end{document}